\documentclass[
preprintnumbers,tightlines,
amsmath,amssymb,aps,twocolumn,nofootinbib,nobibnotes]{revtex4}
\usepackage[colorlinks=true, pdfstartview=FitV, linkcolor=green, citecolor=red, urlcolor=magenta]{hyperref}
\usepackage{latexsym}
\usepackage{graphicx}
\usepackage{physics}
\usepackage{hyperref}
\usepackage{color}
\usepackage[normalem]{ulem}
\usepackage{epsfig}
\usepackage{comment}
\usepackage{caption}
\usepackage{subcaption}
\usepackage{bm}
\def \bre{the Barrow entropy}

\begin{document}


\title{Thermodynamic Topology and Photon Spheres Analysis of Black Holes in Brane-World: Insights from Barrow Entropy}

\author{Usman Zafar $^{1}$\footnote{s2471001@ipc.fukushima-u.ac.jp,~zafarusman494@gmail.com}}
\author{Abdul Jawad $^{2,3}$
\footnote{jawadab181@yahoo.com; abduljawad@cuilahore.edu.pk}}
\author{Kazuharu Bamba $^{1}$\footnote{bamba@sss.fukushima-u.ac.jp}}
\author{Mohammad Ali S. Afshar $^{4,5,6}$\footnote {m.a.s.afshar@gmail.com}}
\author{Mohammad Reza Alipour$^{4,5}$\footnote {mr.alipour@stu.umz.ac.ir}}
\author{Saeed Noori Gashti $^5$\footnote {saeed.noorigashti@stu.umz.ac.ir; saeed.noorigashti70@gmail.com}}
\author{Jafar Sadeghi $^{4,5,6}$\footnote{~pouriya@ipm.ir}}
\address{$^1$Faculty of Symbiotic Systems Science, Fukushima University,
Fukushima 960-1296, Japan.}
\address{$^2$~Institute for Theoretical Physics and Cosmology,\\ Zhejiang
University of Technology, Hangzhou 310023, China}
\address{$^3$ Department of Mathematics, COMSATS University Islamabad,
Lahore-Campus, Lahore-54000, Pakistan.}
\address{$^4$Department of Physics, Faculty of Basic
Sciences, University of Mazandaran, P. O. Box 47416-95447, Babolsar, Iran.}
\address{$^5$School of Physics, Damghan University, P. O. Box 3671641167, Damghan, Iran.}
\address{$^6$ Canadian Quantum Research Center, 204-3002 32 Ave Vernon, BC V1T 2L7, Canada}

\date{\today}

\begin{abstract}
We explore the thermodynamics and geothermodynamics of black holes with \bre~in a brane-world scenario, where the horizon geometry of the black hole is regarded as a fractal structure. Our analysis reveals the behavior of heat capacity, identifying both bound and divergence points. For the Bekenstein-Hawking entropy, the divergence point exhibits smooth behavior, indicating no phase transition. In contrast, we observe divergence with Barrow entropy as the deformation parameter increases, confirming the presence of a zero point in heat capacity through various thermodynamic geometry formalisms. Additionally, we delve into thermodynamic topology, detailing the classification of black holes in the brane-world context and comparing their characteristics determined from the Bekenstein-Hawking and the Barrow entropy. Notably, fixing the deformation and cosmological parameters results in a topological charge $-1$ predominately by the dark matter parameter, which remains unaffected despite variations in other parameters. In the dS model, the cosmological horizon prevents stable photon spheres, making topological charges of $0$ and $+1$ unattainable. Incremental increases in the cosmological parameter reduce the dark matter parameter-dominated region.

\end{abstract}

\maketitle


\section{Introduction}
The study of black holes (BHs) is a fascinating and dynamic field that provides valuables information into the impacts of gravity. Over the past years,  brane-world (BW) scenarios have provided the framework to combine gravity with other fundamental forces, which is an intriguing topic in theoretical physics. In these models, the observable universe is confined to a four-dimensional hypersurface, referred to as a brane, which exists within a higher-dimensional space known as the bulk. It is known that BW models have yet to be studied in recent years, but these models offer significant physical motivation,
such as hierarchy Problems and forces unification. In Ref.~\cite{Randall:1999ee}, Randall and Sundrum (RS-I) established a well-known model in the BW context that resolves hierarchy problems with two branes; however, the RS-II model assumes a single brane with positive tension \cite{Randall:1999vf}.\\
\indent Similarly, some BW models addressed the cosmological constant problem and provided various insights regarding dark matter and energy. For instance, the impacts associated with dark matter could be comprehended using gravity leaking from the extra dimension \cite{Heydar-Fard:2007ahl,Lue:2002sw,Maartens:2010ar}.
Furthermore, in the BW context, numerous studies considered the black strings solution rather than a simple extension of the five-dimensional spacetime metric (for more details in terms of BW scenarios like their thermodynamic, geometry, effect of dimensional crossover, and their shapes, check \cite{Heydar-Fard:2007ahl,Lue:2002sw,Maartens:2010ar,Chamblin:1999by,Emparan:1999wa,Kudoh:2003xz,Abdolrahimi:2012qi,Heydarzade:2017xbb,Shahzad:2021nqt,Harko:2004ui, Dadhich:2000am,Casadio:2003vk,Hollands:2012sf,Li:2015vqa,Jawad:2023ypn}). For instance, the black string solution is more straightforward than the localized five-dimensional BH. It maintains the symmetry of the four-dimensional BH along the extra dimension, making it easier to investigate theoretically. Similarly, one can obtain the black string solution analytically compared to the fully localized five-dimensional BHs, which require numerical techniques. \\
\indent In physics, BH thermodynamics is a specialized area aimed at examining and finding evidence of the alignment of thermodynamic laws with the general behavior of BHs \cite{f,g,h}. Researchers can better understand gravity’s structure and behavioral nature by exploring BH thermodynamics. This area of study has significantly enhanced our comprehension of quantum gravity \cite{b,d}, in much the same way that the investigation of black-body radiation paved the way for quantum mechanics. A significant outcome of this work is the establishment of the holographic principle, proposing that all the information inside a given spatial volume can be described by a boundary theory, highlighting a fundamental connection between quantum mechanics and gravity \cite{e, i, j}.\\
\indent Furthermore, when the size of the BH decreases, its temperature rises, leading to a complete runaway thermal mechanism as described in Refs.~\cite{Shahzad:2021nqt,Hawking:1974rv,Hawking:1975vcx,Bekenstein:1973ur,Page:1976df}. One of the most crucial properties of BHs is their thermal stability, which gives us valuable insights regarding phase transition and critical points \cite{Hendi:2018sbe,Sadeghi:2016dvc,Azreg-Ainou:2014twa,Cai:2014znn,Cai:2013qga}. 
For example, the local stability of a BH depends on the sign of its heat capacity: A positive heat capacity indicates stability, meaning small temperature fluctuations do not trigger runaway effects, whereas a negative heat capacity signifies instability \cite{Hawking:1982dh,Davies:1989ey,Kubiznak:2012wp,Mo:2014mba,Santos:2022vet}. Moreover, thermodynamic geometry is an approach for investigating BHs that interprets thermodynamic parameters (like temperature, heat capacity, and entropy) into the geometric structure \cite{Weinhold:1975xej,Ruppeiner:1995zz}. The main idea of this approach is to utilize the thermodynamic characteristics of a system to compute the Ricci curvature scalar within a multi-dimensional space, and this curvature offers crucial insights regarding the system's consistency and phase transition. However, the complex link between entropy, temperature, free energy, and other parameters in BH solutions makes the thermodynamic geometry approach particularly useful in BH thermodynamics, even though it is not restricted to BHs and can be applied to various fields. Furthermore, numerous works have been done employing the thermodynamic geometry formalism in various modified entropies (to comprehend better the utilization of the modified entropies in the geothermodynamic formalisms, see Refs.~\cite{Jawad:2023sul, Soroushfar:2020wch}). Recently, Barrow proposed a new BH entropy in Ref.~\cite{Barrow:2020tzx}. The primary motivation was to incorporate possible quantum gravitational influences that could lead to a fractal or irregular structure in a BH's event horizon at microscopic levels, rather than maintaining perfect smoothness. The modification in the entropy formula was intended to explore the possible impacts of quantum gravity on BH thermodynamics and informational aspects. In this work, we consider the \bre~to examine the thermodynamic stability and phase transition of BW BH. By utilizing \bre~in BW BHs, we can analyze the modifications in their thermodynamics caused by extra dimensions, potentially deepening insights into quantum gravity, which might provide a more comprehensive understanding of how quantum gravitational corrections, extra dimensions, and BH thermodynamics are interconnected.  \\
\indent Recently, groundbreaking methods have emerged for studying and determining the phase transition in BH thermodynamics. Among these, the topological method stands out, as outlined in several studies \cite{18a,19a,20a,21a}. This approach offers a novel framework to investigate the physics of BHs and proposes a scheme for classifying BHs according to their topological charges. It has been proposed that topological analysis can be applied to analyze critical phenomena in BH thermodynamics. Their analysis revealed two separate categories of critical points, overlooked in earlier studies, referred to as novel and conventional critical points, each linked to a particular topological charge \cite{18a, 19a}. This topological classification has opened new avenues for understanding the phase transitions and stability of BHs. By examining the topological properties, scientists can gain insights into the fundamental nature of spacetime and the behavior of BHs under various conditions. This approach also helps identify universal properties of BHs that are independent of the specific details of spacetime geometry \cite{18a, 19a}.\\
\indent Duan's topological current $\phi$-mapping theory is crucial for adopting a topological perspective in thermodynamics. In Refs.~\cite{18a, 19a}, it has been suggested that there are two separate approaches for examining topological thermodynamics,  focusing on the generalized free energy and temperature function. The temperature approach breaks down the function of the temperature by eradicating the pressure and incorporating auxiliary and topological parameters expressed as $1/\sin\Theta$. These ideas serve as the foundation for the potential's further construction. In the other technique, the generalized free energy function treats BHs as defects within the thermodynamic parameter space. Their solutions is investigated by utilizing the generalized off-shell free energy, where the instability and stability of BH solutions interpreted by negative and positive winding numbers, respectively (To better understand the thermodynamic topology by utilizing generalized free energy in order to obtain topological charge or winding number, check Refs.~\cite{18a,19a,20a,21a,23a,24a,25a,26a,28a,30a,31a,33a,34a,35a,38a,39a,40a,41a,42a,43a,44a,46a,Bhattacharya:2024bjp,Yasir:2024wir,5000,5001,5002,5003,5004,5005,5006,Wu:2022whe,Zhu:2024jhw,Wu:2024asq,Chen:2025nto,Ai:2025vno,Wu:2023xpq,Wu:2023fcw,Chen:2024atr,Liu:2025iyl,Chen:2025fse,Wu:2025wpz,Rani:2025mip}). \\
\indent In this study, we will utilize the Helmholtz free energy method to analyze the selected model under two different structural forms of entropy: the Bekenstein-Hawking and \bre~\cite{Bekenstein:1973ur,Barrow:2020tzx,7010,7011,7012,7013,7014,7015,7016}. In this work, we will demonstrate how applying topological thermodynamic charge analysis to these entropies can yield profound insights into the BH's thermodynamic properties. In theoretical physics, the study of topological photon spheres in BHs explores regions where the gravitational field is intense enough to trap particles of light (photons) in circular orbits. This region is very significant for examining BH characteristics, notably their shadow and the phenomena of light bending \cite{49a, 50a}. Researchers employ topological methods to examine the properties and the conduct of photon spheres. These techniques elucidates the stability behavior along with the photon sphere's dynamics and its interactions with other physical effects in the BH's vicinity. This article aims to describe a unique type of BH known as the BW BH.  We choose \bre~ to compare the case of the Bekenstein-Hawking entropy and find new insights into BH thermodynamics; we consider the case of \bre~ because it comprises the consideration of the fractal-like structure of the horizon as detailed in Refs.~\cite{Barrow:2020tzx, Capozziello:2025axh}. Therefore, we will explicitly use \bre~ to explore this BH from the perspective of thermodynamics, thermodynamic geometry, and thermodynamic topology. Our aims include identifying the topological classifications and charges of these BHs as well as determining the probability of their associated photon spheres.\\ 
\indent Although thermodynamic geometry and topological approaches utilizing \bre~have been studied for numerous BH spacetimes, our results demonstrate that the braneworld framework introduces distinct, model-specific effects. The heat-capacity divergences emerge uniquely for \bre~ but remain absent for Bekenstein-Hawking entropy, providing clear evidence of a non-trivial interaction between quantum fractal horizon corrections and bulk-origin cosmological and dark-matter parameters. Moreover, despite the universal agreement between heat capacity zero point (or divergence) and geothermodynamic metric's curvature scaler singularities, the thermodynamic topological charge is primarily dictated by the dark-matter parameter $\beta$, reflecting a fundamental property of the BW BH rather than a generic effect of entropy modifications. This work is arranged in the manner described hereafter. In
\texttt{Sec.~\ref{Sec-2}}, we discuss the solution of the brane word BH and incorporate the thermodynamic parameters, which are crucial to studying the thermodynamic features of BHs.  In
\texttt{Sec.~\ref{Sec-3}}, we investigated the BW BH's stability and discussed the bound and divergence points using \bre. Further, we have investigated the influence of dark matter, cosmological, and deformation parameters on the thermodynamics of BW BHs. Moreover,  in \texttt{Sec.~\ref{Sec-4}}, we study the geothermodynamics of the BE BH
by using \bre. We introduce the idea of topological thermodynamics and thoroughly describe the mathematical formalisms of our analysis in \texttt{Sec.~\ref{Sec-5}}. This involves categorizing the BW BH with respect to its topological properties, determining its topological charges, and addressing the significance of these classifications. This section also focuses on the comparative analysis of the thermodynamic properties computed from the Bekenstein-Hawking and \bre frameworks, highlighting their major differences and similarities.
In \texttt{Sec.~\ref{Sec-6}}, we discuss the BH's photon sphere and investigate the role of free parameters in this context.
In \texttt{Sec.~\ref{Sec-7}}, we provide the concluding remarks of our investigation and discuss the possible extension of this work within BH thermodynamics and topological analysis.

\section{Brane-World Black Hole: A Brief Discussion} \label{Sec-2}

In this section, we have discussed the BW BH solution provided in Ref.~\cite{Heydar-Fard:2007ahl}. The Einstein field equation for the BW BH comprises the bulk effect (higher dimensional), extra-dimensional geometry, and the cosmological constant, and it is obtained by making some adjustments as discussed in Ref.~\cite{Heydar-Fard:2007ahl}. Thereby, the final form of the field equation in a vacuum field is given by
\begin{eqnarray}\label{EFE}
G_{ab}=Q_{ab}-\epsilon_{ab}-\Lambda g_{ab}\,,
\end{eqnarray}
where  $\epsilon_{ab}$ is a symmetric tensor associating to the bulk curvature, $\Lambda$ is the cosmological constant and $g_{ab}$ represents the metric tensor. Furthermore, the term $Q_{ab}$ is a geometrical correction term that comes from the bulk effect in the context of the BW. However,  as we know that $\epsilon_{ab}$ is a symmetric tensor, and by employing the Bianchi identities (a mathematical statement), which are used for conservation equations,  then we have the following relation
\begin{eqnarray}\label{SMT}
\epsilon^{ab}=0\,.
\end{eqnarray}

By utilizing Eq.~(\ref{SMT}), it is a straightforward approach to solve the system of equations in the vacuum field for the BW. In addition,  by assuming that cosmological constant $(\Lambda)$ and $\epsilon_{ab}$ are equal to zero in Eq.~(\ref{EFE}), it yields 
\begin{eqnarray}\label{VFE}
G_{ab}=Q_{ab}\,.
\end{eqnarray}
The reason for putting  $\Lambda=0$ is that it permits us to study other aspects of the BW model without including the complexity of the cosmological constant, and also, one can easily compare this solution to the classical solution, like the Schwarzschild BH.  Furthermore, as we know, $Q_{ab}$ is the geometric term, and it is important to comprehend how the bulk geometry impacts the physics of the BW. In addition, it involves the extrinsic curvature term, which describes how the Brane is curved within the bulk. Thus, mathematically, it takes the given form
\begin{eqnarray}\nonumber
Q_{ab}&=&\big(K K_{ab}-K_{a\alpha}K_{b}^{\alpha}\big)+\displaystyle{\frac{1}{2}g_{ab}}\big(K_{\alpha\beta}K^{\alpha\beta} \\\label{ETC}&-&K^{2}\big)\,,
\end{eqnarray}
where $K_{ab}$  represented the extrinsic curvature tensor and $K$ is its trace.   Moreover, the combination of $K_{\alpha\beta}K^{\alpha\beta} -K^{2}$ in the right-hand side of Eq.~(\ref{ETC}) is similar to the Gauss-Coadazzi equation, which links the extrinsic curvature in bulk to the intrinsic curvature of the Brane. Similarly, the first term of the above equation incorporates the interaction between the component of $K_{ab}$ and its trace $K$.
Thereby, the line element for the BW BH is given as 
\begin{eqnarray}\label{LINE}
ds^{2}=-A(r)dt^{2}+\frac{1}{A(r)}dr^{2}+r^{2}d\Omega^{2}\,,
\end{eqnarray}
where $d\Omega^{2}=d\theta^2+\sin(\theta)^{2}d\phi^2$. According to Refs.~\cite{Heydar-Fard:2007ahl, Heydarzade:2017xbb, Shahzad:2021nqt}, the corresponding metric coefficient for this line element takes the following shape
\begin{eqnarray}\label{METF}
A(r)=1-\alpha^{2}r^{2}-2\alpha \ \beta \ r-\beta^{2}-\frac{2M}{r}\,,
\end{eqnarray}
where $\alpha$, $\beta$, and $M$ represent the cosmological parameter, dark matter parameter, and the mass of the BH, respectively. Here, we mentioned that the linear term in Eq.~(\ref{METF}) represents the consequences of curvature in higher dimensions. We can remove this term by making some adjustments, like coordinate transformation or by adjusting the gauge conditions (for further details in this context, check Refs.~\cite{Harko:2004ui, Dadhich:2000am, Maartens:2010ar}). In addition, BH stability could be modified due to the linear term's effect on the horizon's configuration, and its removal could obscure an essential physical property that differentiates BW BHs from standard BH theories. Thermodynamic attributes of the BH, which play a crucial role in its dynamic equilibrium, are influenced by the incorporation of the linear term \cite{Harko:2004ui, Dadhich:2000am, Maartens:2010ar,Casadio:2003vk,Hollands:2012sf,Li:2015vqa}. Furthermore, if we put $\alpha$ and $\beta$ equal to zero, one can easily obtain the Schwarzschild BH solution, one of the first BH solutions. So, we focus on exploring the BW BH's \footnote{We mentioned here the key distinction between the Schwarzschild BH with cosmological constant and BW BH is that BW BH arises due to the higher dimensional theories of gravity. For example, the parameters $\alpha$ and $\beta$ do not originate from a straightforward cosmological constant but instead arise due to bulk-brane interactions. Due to extra-dimensional effects, the metric function gains a linear term and a quadratic term, which resemble cosmological influences but originate from extra-dimensional physics rather than the usual cosmological constant $\Lambda$  \cite{Chamblin:1999by,Maartens:2010ar, Perlick:2018iye,Svarc:2018coe}. } thermodynamic properties with respect to \bre. 

Here, we want to highlight our motivation for departing from the standard entropy formulation (Bekenstein-Hawking formalism) to \bre. While BH entropy is generally linked to its horizon area, the inclusion of quantum gravitational effects challenges this classical concept. For example, at the quantum scale, a BH's structure is very complex and intricate, allowing us to investigate the thermodynamic properties of the BH by modifying the conventional entropy. In Ref. \cite{Barrow:2020tzx}, this issue of intricate structure (fractal structure) has been addressed by modifying the association between entropy and horizon area, so that the horizon area of the BH changes to $r^{\delta/2+1}$, then the entropy relevant to this modified horizon surface is also modified, which takes the given shape 
\begin{eqnarray}\label{BREF}
S=(A/A_{pl})^{1+\delta/2}\,,
\end{eqnarray}
where $A_{pl}$, $\delta$, and $A$ are the Plank's area, deformation parameter, and area of the horizon of the BH, respectively. It is worth mentioning that the $\delta$ is dimensionless; it indicates the intricacy or complexity of the structure of the BH's surface area, and its range lies in the interval of $[0,~1]$. However, after doing some adjustments in Eq.~(\ref{BREF}), we can determine the expression for \bre, which is given as 
\begin{eqnarray}\label{BRE}
S=(\pi \ r^{2}_{e})^{\frac{\delta+2}{2}}\,,
\end{eqnarray}
where $r_{e}$ is the horizon radius. Furthermore, it can be observed that $\delta=0$ corresponds to the smooth horizon surface, which is equivalent to the Bekenstein-Hawking entropy, whereas $\delta=1$ reflects the most intricate structure. The mass of the BH $M$ is obtained by substituting Eq.~(\ref{METF}) into the expression $A(r_{e})=0$, which is given as
\begin{eqnarray}\label{MM}
M&=&-\frac{1}{2} r_{e} \left(\beta ^2+\alpha ^2 r_{e}^{2}-1+2 \alpha  \beta  r_{e}\right)\,,
\end{eqnarray}
By utilizing Eqs.~(\ref{BRE}) and (\ref{MM}), the relation for the BW BH's mass can be described with respect to \bre, as follows
\begin{eqnarray}\nonumber
M&=&-\frac{S^{\frac{1}{\delta +2}}}{2 \pi ^{3/2}} \bigg\{\pi  \left(\beta ^2-1\right)+\alpha ^2 S^{\frac{2}{\delta +2}}+2 \sqrt{\pi } \alpha  \beta \\\label{MBRE}&\times& S^{\frac{1}{\delta +2}}\bigg\}\,.
\end{eqnarray}
It is worth mentioning that we define the BH mass $M$ employing the condition $A(r_{e})=0$, which identifies the event horizon, as presented in Eq.~\eqref{MM} \cite{Heydar-Fard:2007ahl,Maartens:2010ar,Shahzad:2021nqt}. We explicitly check that this mass satisfies thermodynamic consistency within the Barrow entropy formalisms. By employing Eq.~\eqref{MBRE}, the conjugate temperature corresponding to the Barrow entropy can be determined. This relationship is explicitly given by
\begin{eqnarray}\nonumber
T&=&\frac{\partial{M}}{\partial{S}}=\frac{S^{\frac{1-\delta}{\delta +2}}}{2 \pi ^{3/2} (\delta +2)} \bigg\{\pi  \left(\beta ^2-1\right)+3 \alpha ^2 S^{\frac{2}{\delta +2}}\\\label{TBRE}&+&4 \sqrt{\pi } \alpha  \beta  S^{\frac{1}{\delta +2}}\bigg\}\,.
\end{eqnarray}
We observe that the mass function is consistent with the first law of thermodynamics under Barrow entropy, confirming its validity as the thermodynamic mass which acts as the internal energy as given in Refs.~\cite{Kastor:2009wy,Barrow:2020tzx,Capozziello:2025axh,Zafar:2025sxl}. If we put $\delta=0$, one can easily determine the Hawking temperature from Eq.~\eqref{TBRE}. Consequently, the formulation maintains internal consistency from both the geometric and thermodynamic perspectives.
\section{Thermal Stability of Brane-World BH through \bre}  \label{Sec-3}

In this section, we analyze the stability and phase transition of the BW BH by utilizing \bre, and to analyze the thermal stability, we employ the heat capacity $(C)$, which is one of the approaches to explore the BH's stability. Furthermore, the stability of the BH is linked with the negative and positive values of $C$; for instance, if $C < 0 $, then it signifies the instability of the BH, while the stable phase is linked with the $C>0$. Moreover, another interesting aspect of heat capacity $C$ is that it also provides an interpretation of phase transition through its divergence as described in Ref.~\cite{Hendi:2018sbe}. The expression of heat capacity $C$ is represented  as
\begin{eqnarray}\label{CBH}
C=T\left(\frac{\partial M}{ \partial S} \right)=\frac{T}{\partial^{2} {M}/\partial{S^2}}\,.
\end{eqnarray}
Thereby, the heat capacity is computed to investigate the thermal stability and phase transition of the BW BH  by employing Eq.~(\ref{TBRE}) in Eq.~(\ref{CBH}), which is given by
\begin{eqnarray}\nonumber
C&=&\bigg\{ (\delta +2) S \bigg[-\pi  \left(\beta ^2-1\right)-3 \alpha ^2 S^{\frac{2}{\delta +2}}-4 \sqrt{\pi } \alpha  \beta\\\nonumber&\times&  S^{\frac{1}{\delta +2}}\bigg]\bigg\}\times\bigg\{\pi  \left(\beta ^2-1\right) (\delta +1)+3 \alpha ^2 (\delta -1) S^{\frac{2}{\delta +2}}\\\label{CBRE}&+&4 \sqrt{\pi } \alpha  \beta  \delta  S^{\frac{1}{\delta +2}}\bigg\}^{-1}\,.
\end{eqnarray}
\begin{figure*}[ht]
     \begin{subfigure}{0.4\textwidth}
         \includegraphics[width=\textwidth]{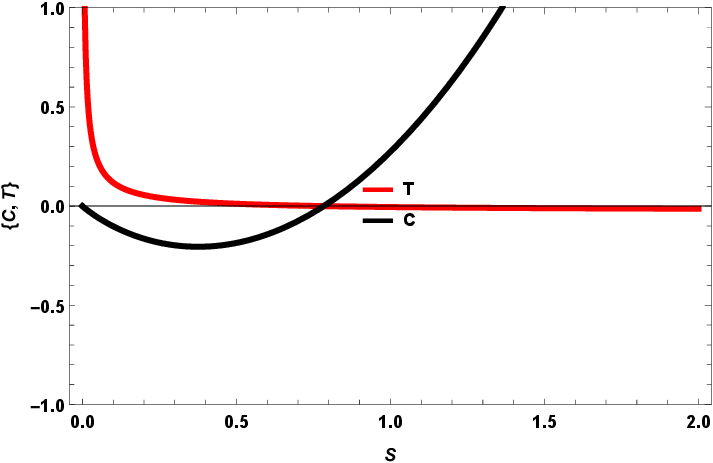}
         \caption{$\alpha=0.3,\beta=0.7$}
         \label{C, T versus S}
     \end{subfigure}
     \begin{subfigure}{0.4\textwidth}
         \includegraphics[width=\textwidth]{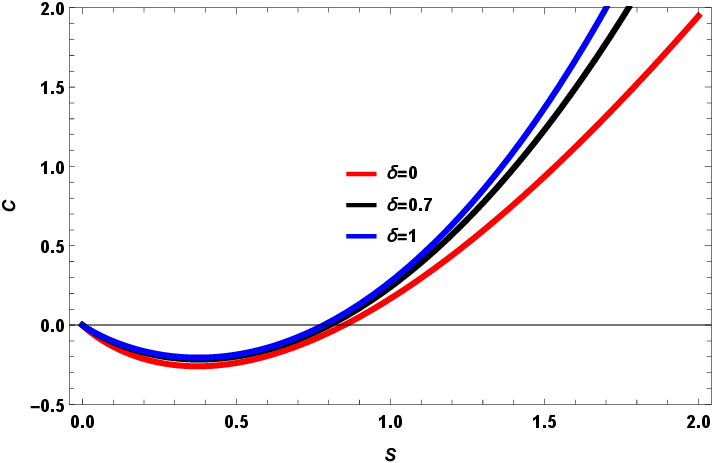}
         \caption{$\alpha=0.3,\beta=0.7$}
         \label{Cgraph}
     \end{subfigure}
        \caption{\raggedright Heat capacity $C$ and temperature $T$ as the function of \bre. In Fig.~\ref{CTGRAPH}(\subref{C, T versus S}), we have plotted $T$ (red curve) and heat capacity $C$ (black curve) by inserting $\alpha=~0.3,~\beta=~0.7$ and deformation parameter $\delta=~1$. In Fig.~\ref{CTGRAPH}(\subref{Cgraph}), heat capacity is presented by inserting $\delta=0$ (red curve), $\delta=0.5$ (black curve), $\delta=1$ (blue curve).}
        \label{CTGRAPH}
\end{figure*}

 Furthermore, examining Eq.~(\ref{CBH}) reveals that the non-physical and physical BH solutions can be distinguished by studying the roots of the temperature. Therefore, the physical solution of BHs is associated with the positive value of the temperature, while the non-physical solution corresponds to the negative value of the temperature. However, exploring the bound points and the phase transition in the heat capacity for the BW BH is crucial. As we know, bound points of heat capacity can be obtained from the numerator of Eq.~(\ref{CBH}), while for divergence, one can utilize the denominator of Eq.~(\ref{CBH}) as discussed in Ref.~\cite{Shahzad:2021nqt}. Therefore, in this work, we have used Eq.~(\ref{CBRE}) to obtain the bound point and divergence of the heat capacity in the form of \bre.  

Moreover, we have extracted two types of roots from Eq.~(\ref{CBRE}) numerator and a point of divergence from the denominator. Here, our goal is to explore how the deformation parameter of \bre~influences the thermal stability of the BW BH. We have graphically presented the behavior of heat capacity (black curve) and temperature (red curve) by inserting $\alpha=0.3~\beta=0.7$ and $\delta=1$ in Fig.~\ref{CTGRAPH}(\subref{C, T versus S}). While various trajectories of heat capacity are presented by putting $\delta=0$ (red curve), $\delta=0.7$ (black curve), and $\delta=1$ (blue curve) for $\alpha=0.3$, and $\beta=0.7$ in Fig.~\ref{CTGRAPH}(\subref{Cgraph}). Thus, it can be noticed that heat capacity has a zero point (ZP) at $S_{0}=0.7837$, and the location of this point can be varied by choosing the various values of $\delta$. For instance, the ZP corresponding to the Hawking-Bekenstein $\delta=0$ (red curve) is $0.850$, and in the case of most intricate structures $\delta=1$ (blue curve), it is $0.7837$.  The heat capacity's ZP indicates the transition from an unstable to a stable phase, or vice versa. In addition, as explained in Ref.~\cite{Shahzad:2021nqt}, we can use the above reasoning to examine the roots and points of divergence derived from Eq.~(\ref{CBRE}), which yields
\begin{eqnarray}\nonumber
&S_{\mathrm{Bp}}&=\left(\frac{-\sqrt{\pi } \sqrt{\beta ^2+3}-2 \sqrt{\pi } \beta }{3 \alpha }\right)^{\delta +2}\,,\\\label{Roots}& S_{\mathrm{Bp}}&=\left(\frac{\sqrt{\pi } \sqrt{\beta ^2+3}-2 \sqrt{\pi } \beta }{3 \alpha }\right)^{\delta +2}\,.
\end{eqnarray}
Here, we mentioned that $\mathrm{Bp}$ are the bound points in terms of \bre. Furthermore, if we put $\delta=0$, we can obtain the bound points of heat capacity for the Bekenstein-Hawking entropy, and one can also analytically see the effect of \bre~on the bound point of heat capacity. Now, we discuss the points of divergence by using the denominator of Eq.~(\ref{CBRE}), which is given by
\begin{eqnarray}\nonumber
S_{\mathrm{Dp}}&=&\bigg\{-\bigg[\sqrt{\pi } \sqrt{\beta ^2 \left(\delta ^2+3\right)+3 \left(\delta ^2-1\right)}+2 \sqrt{\pi } \beta  \\\label{DP}&\times&\delta \bigg]\times\bigg[3 \alpha  (\delta -1)\bigg]^{-1}\bigg\}^{\delta +2}\,.
\end{eqnarray}

Moreover, we obtained two divergence points from the denominator of Eq.~(\ref{CBRE}), but we neglected the negative one and considered only the positive, which is mentioned in Eq.~(\ref{DP}). Let us mention here that we cannot put $\delta=1$ for the divergence point because it becomes undefined (infinite). Furthermore, one can quickly notice the impact of \bre~ in Eq.~(\ref{DP}), and also, by plugging $\delta=0$, we can obtain the divergence point in the form of the Bekenstein-Hawking entropy. Therefore, we can use the given range $0 \leq \delta < 1$ for divergency.

\begin{table}[t]
\caption{\label{table:1}
\raggedright Summary of the bound and divergency points by keeping dark matter parameter $\beta=~-1.4$ constant (upper panel) and dark matter parameter $\alpha=~0.7$  constant (lower panel).}
\begin{ruledtabular}
\begin{tabular}{ccccc}
$\alpha$ & $\delta$ & Small root & Large root & Divergence \\
\hline
 0.7 & 0 & 0.233808 & 18.0031 & 2.05165 \\
 0.7 & 0.3 & 0.188014 & 27.7747 & 1.48122 \\
 0.7 & 0.7 & 0.140592 & 49.513 & 0.918559 \\
 0.7 & 1 & 0.113055 & 76.3873 & \text{Indeterminate} \\
 0.9 & 0 & 0.14144 & 10.8908 & 1.24112 \\
 0.9 & 0.3 & 0.105477 & 15.5818 & 0.830976 \\
 0.9 & 0.7 & 0.0713298 & 25.1205 & 0.466034 \\
 0.9 & 1 & 0.0531933 & 35.9408 & \text{Indeterminate} \\
\hline
\end{tabular}
\begin{tabular}{ccccc}
$\beta$ & $\delta$ & Small root & Large root & Divergence \\
\hline
 -1.1 & 0 & 0.0156402 & 12.8784 & 0.448799 \\
 -1.1 & 0.3 & 0.00838257 & 18.8948 & 0.132512 \\
 -1.1 & 0.7 & 0.00364943 & 31.5003 & 0.0334709 \\
 -1.1 & 1 & 0.00195597 & 46.2162 & \text{Indeterminate} \\
 -1.4 & 0 & 0.233808 & 18.0031 & 2.05165 \\
 -1.4 & 0.3 & 0.188014 & 27.7747 & 1.48122 \\
 -1.4 & 0.7 & 0.140592 & 49.513 & 0.918559 \\
 -1.4 & 1 & 0.113055 & 76.3873 & \text{Indeterminate} \\
\hline
\end{tabular}
\end{ruledtabular}
\end{table}

\begin{figure*}
     \begin{subfigure}[b]{0.45\textwidth}
         \includegraphics[width=\textwidth]{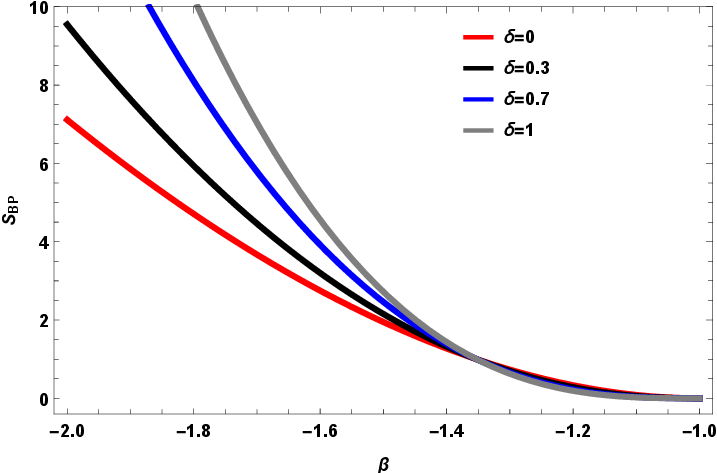}
         \caption{Small root}
         \label{BP3}
     \end{subfigure}
     \begin{subfigure}[b]{0.45\textwidth}
         \includegraphics[width=\textwidth]{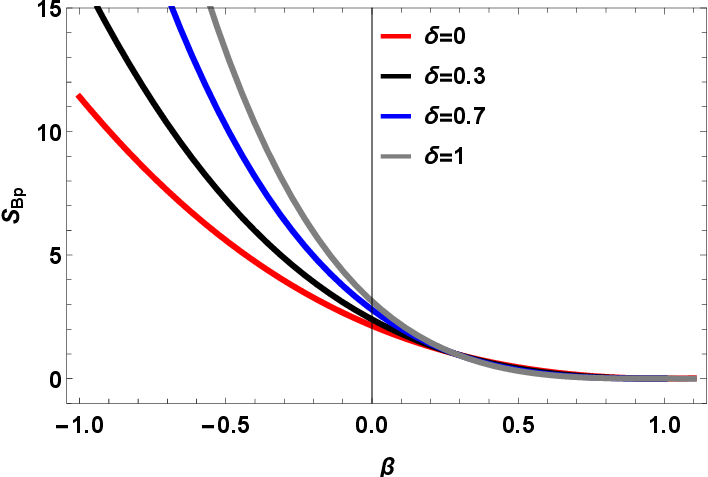}
         \caption{Large root}
         \label{BP4}
     \end{subfigure}
      \begin{subfigure}[b]{0.45\textwidth}
         \includegraphics[width=\textwidth]{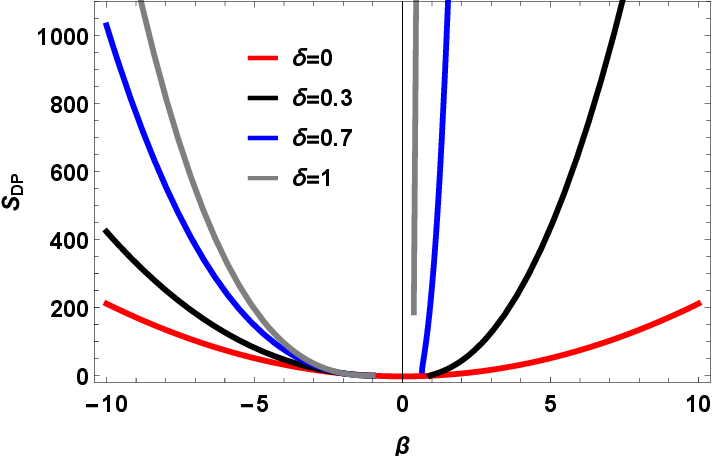}
         \caption{Divergency}
         \label{DP2}
     \end{subfigure}
        \caption{\raggedright Small root (left), large root (middle), and divergence (right) in terms of dark matter parameter $\beta$. In Figs.~\ref{BDGRAPHB}(\subref{BP3}),~\ref{BDGRAPHB}(\subref{BP4}) and \ref{BDGRAPHB}(\subref{DP2}), the trajectories correspond to the different values of deformation parameters such as $\delta=0$ (red curve), $\delta=0.3$ (black curve), $\delta=0.7$ (blue curve) and $\delta=1$ (grey curve).}
        \label{BDGRAPHB}
\end{figure*}

Furthermore, we investigate the effects on the bound points and divergence points of heat capacity in more detail for various values of $\delta,~\alpha$, and $\beta$ by employing Eq.~(\ref{Roots}) and (\ref{DP}), described in  Table~\ref{table:1}. We have fixed cosmological parameter $\alpha$ and dark matter parameter $\beta$ in these tables by using different values of $\delta$ for instance, we have fixed dark matter parameter $\beta=~-1.4$ in Table~\ref{table:1} (upper panel), while for Table~\ref{table:1} (lower panel), we have fixed cosmological parameter $\alpha=0.7$. Here, we mentioned that the values of $\alpha$ and $\beta$ are obtained as discussed in Ref.~\cite{Shahzad:2021nqt}. In our analysis, one can easily observe the influence of $\delta$ on the thermodynamics of these BHs. If we look at Table~\ref{table:1} (lower and upper panel), it can be noticed that when the deformation parameter $\delta$ rises, the small root and the divergence point decrease while the large root increases rapidly.\\ 
\indent
Moreover, in Fig.~\ref{BDGRAPHB}, we have graphically presented the behavior of bound points and divergence concerning the dark matter parameter $\beta$ by using  $\delta=0$ (red curve), $\delta=0.3$ (black curve), $\delta=0.7$ (blue curve) and $\delta=1$\footnote{For divergency, we have inserted the $\delta=0.9$ instead of $\delta=1$ in Fig.~\ref{BDGRAPHB}(\subref{DP2}) because at $\delta=1$ divergency root becomes infinite.} (grey curve). From Fig.~\ref{BDGRAPHB}, it is apparent that stability rises as the dark matter parameter $\beta$ increases and decreases with lower values, indicating that inconsistency increases for the small values of dark matter parameter $\beta$. Furthermore, the divergence shown in Fig.~\ref{BDGRAPHB}(\subref{DP2}) is absent in the Bekenstein-Hawking entropy case, demonstrating that it is not an inherent property of the BH. Rather than being generic, the observed behavior arises from the non-additive \bre~ modifications, which significantly alter the entropy-temperature correspondence and play a dynamic role when combined with the BW dark-matter parameter $\beta$. Furthermore, these divergences in the case of \bre~ are verified using thermodynamic geometry. From both Fig.~\ref{BDGRAPHB}, it can be deduced that the zero point (or divergence) of heat capacity $C$ can be avoided for large values of $\alpha$ and $\beta$. 

\section{Thermodynamic Geometry of Brane-World Black Hole Through \bre}  \label{Sec-4}

This section presents various geothermodynamic formalisms by employing \bre~ and discusses the coincidence of divergence of Ricci scalar with ZPs of heat capacity. First, we will discuss the Weinhold metric \cite{Weinhold:1975xej} because it is considered one of the initial and basic formalisms in thermodynamic geometry. Here, we mentioned that our mass is the function of $M(S,~\alpha,~\beta)$ instead of $M(S,~l,~Q)$ because, in the BW BH scenario, we have cosmological parameter $\alpha$ and dark matter parameter $\beta$. Thereby, the Weinhold metric formalism in the case of the Brane-word BH can be expressed as 
\begin{eqnarray}\label{WMF}
g_{\nu\omega}^{\mathrm{W}}=\partial_{\nu}\partial_{\omega} M(S,~\alpha,~\beta)\,,
\end{eqnarray}
and the line element, according to the above-mentioned metric formalism, can be written as 
\begin{eqnarray}\nonumber
ds^{2}_{\mathrm{W}}&=&M_{SS}dS^{2}+M_{{\alpha}{\alpha}}d{\alpha}^{2}+M_{{\beta}{\beta}}d{\beta}^{2}+2M_{{\alpha}{\beta}}d{\alpha}d{\beta}\\\label{GMLE}&+&2M_{{S}{\beta}}d{S}d{\beta}+2M_{{\alpha}{S}}d{\alpha}d{S}\,.
\end{eqnarray}
One can write the line element of the Weinhold formalism given in Eq.~(\ref{GMLE}) in the form of a metric, which takes the given shape as
\begin{equation}\label{M1}
g^{\mathrm{W}}=\begin{pmatrix}
M_{SS} & M_{S\alpha} & M_{S\beta}  \\
M_{\alpha S} & M_{\alpha\alpha} & M_{\alpha \beta} \\ M_{\beta S} & M_{\beta \alpha}
  & M_{\beta \beta} \\
\end{pmatrix}\,.
\end{equation}
It is quite easy to obtain the curvature scalar $R^{w}$ from the above metric formalism, and it is given as 
\begin{eqnarray}\label{WRS}
R^{\mathrm{W}}=-\frac{\pi ^{3/2} S^{-\frac{1}{\delta +2}}}{2 \left(\sqrt{\pi } \beta +\alpha  S^{\frac{1}{\delta +2}}\right)^2}\,.
\end{eqnarray}

\begin{figure*}
\epsfig{file=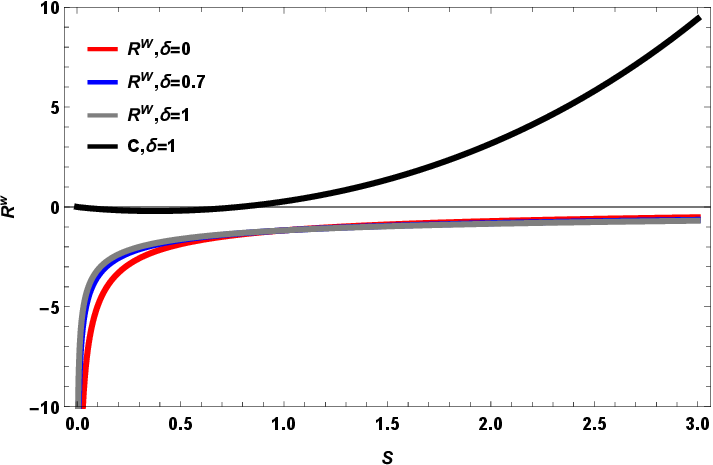,width=.47\linewidth}
\epsfig{file=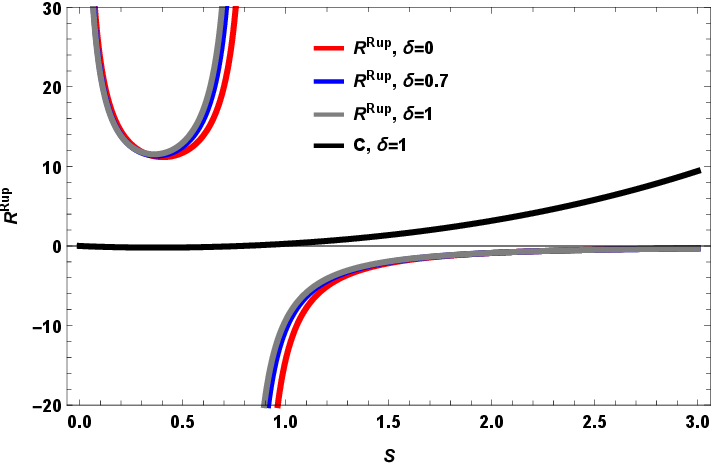,width=.47\linewidth} \caption{\raggedright Plot of $R^{\mathrm{W}},~R^{\mathrm{Rup}}$ and $C$ against $S$ for numerous values of deformation parameter. In the left panel, we presents $C$ (black curve) and $R^{\mathrm{W}}$ is presented in different trajectories for $\delta=0$ (red curve), $\delta=0.7$ (blue curve) and $\delta=1$ (grey curve). In the right panel, we presents $C$ (black curve) and $R^{\mathrm{Rup}}$ is illustrated in various trajectories for $\delta=0$ (red curve), $\delta=0.7$ (blue curve) and $\delta=1$ (grey curve).} \label{figure 4}
\end{figure*}

We have graphically demonstrated Ricci scalar $R^{\mathrm{W}}$ (red curve) and heat capacity (black curve) by inserting $\alpha=0.3,~\beta=0.7$ $\delta=0$ (red curve), $\delta=0.7$ (blue curve) and $\delta=1$ (grey color) in Fig.~\ref{figure 4}  (left panel). It is observed that neither divergence nor the singular point of the Weinhod metric formalism meets the ZP of the heat capacity. The absence of singularity and the nondivergence of the curvature scalar are of no significance.  But if we look at the negative behavior of the $R^{\mathrm{W}}$, it indicates the interaction between the particles of the BW BH is attractive. Therefore, it allows us to shift our investigation of finding the coincidence between the ZP of the heat capacity and divergence or singular points of $R^{\mathrm{W}}$, from the Weinhold to the Ruppeiner formalism \cite{Ruppeiner:1995zz}, which is considered to be a modified form of the Weinhold formalism. The reason we are saying that it is the modified form of the Weinhold metric formalism is that it involves the additional temperature term with the metric of the Weinhold formalism, and it takes the following shape
\begin{eqnarray}\label{RMF}
ds^{2}_{\mathrm{Rup}}=\frac{1}{T}\displaystyle{\left(ds^{2}_{\mathrm{W}}\right)}\,,
\end{eqnarray}
and relevant metric regarding Eq.~(\ref{RMF}) is given by
\begin{equation}\label{M2}
g^{\mathrm{Rup}}=\displaystyle{\frac{1}{T}}\begin{pmatrix}
M_{SS} & M_{S\alpha} & M_{S\beta}  \\
M_{\alpha S} & M_{\alpha\alpha} & M_{\alpha \beta} \\ M_{\beta S} & M_{\beta \alpha}
  & M_{\beta \beta} \\
\end{pmatrix}\,.
\end{equation}

By utilizing Eq.~(\ref{M2}), we obtained the curvature scalar whose denominator is given as  
\begin{eqnarray}\nonumber
denom(R^{\mathrm{Rup}})&=&4 (\delta +2) S \left(\beta +\alpha  \mathcal{B}^{\frac{1}{\delta +2}}\right)^2 \bigg(\beta ^2+3 \alpha ^2 \\\label{RP}&\times& \mathcal{B}^{\frac{2}{\delta +2}}+4 \alpha  \beta  \mathcal{B}^{\frac{1}{\delta +2}}-1\bigg)\,,
\end{eqnarray} 
where $\mathcal{B}=(\pi ^{-\frac{\delta }{2}-1} S)$. In Fig.~\ref{figure 4} (right panel), we have graphically presented the behavior of Ricci curvature scalar $R^{\mathrm{Rup}}$ and $C$ as the functions of \bre~by using $\alpha=0.3,~\beta=0.7$,~$\delta=0$ (red curve), $\delta=0.3$ (black curve), $\delta=0.7$ (blue curve), and $\delta=1$ (grey curve).  As we can observe, the behavior of $R^{\mathrm{Rup}}$ is both negative and positive, indicating that the interaction between the particles of the BHs is both attractive and repulsive. Also, if we examine Fig.~\ref{figure 4}, one can notice that there is divergence in $R^\mathrm{Rup}$, obtained from Eq.~(\ref{RMF}), which also intersects with the ZP of the heat capacity at $S_{0}=0.7837$. 

The intersection of the ZP with the divergence of $R^{\mathrm{Rup}}$ is very significant because it offers valuable insights regarding the stability and phase transition. Varying cosmological parameter $\alpha$, dark matter parameter $\beta$, and deformation parameter $\delta$ might change the location of the ZP of heat capacity and divergence of $R^{\mathrm{Rup}}$. However, the coincidence between the ZP of heat capacity and the divergence of the Ricci curvature scalar will still hold.  Furthermore, it can also be inferred that in the case of the BW BH, one can get more insights from the Ruppeiner metric formalism instead of the Weinhold metric formalism.
\section{Topological interpretation of the Brane-world Black holes}\label{Sec-5}
Thermodynamic topology, though associated with familiar stability diagnostics (like heat capacity and Ruppeiner curvature), offers an independent global perspective by categorizing BH solutions according to conserved topological charges. Local thermodynamic indicators are sensitive to parameter variations, whereas the winding number remains invariant under continuous deformations, capturing the global phase topology of the BH beyond stability analysis. Let us mention that there are primarily two ways to describe the thermodynamic function for thermodynamic topology:  generalized free energy and temperature. In this case, we have employed generalized free energy as our thermodynamic potential because it is a function of the mass and temperature. Thus, by assuming the connection between mass and energy in BHs, it is a straightforward approach to obtain the thermodynamic function, which is given as\cite{25a,31a,38a,40a,44a}
\begin{equation}\label{F1}
\begin{split}
\mathcal{F}=M-\tau^{-1}S\,.
\end{split}
\end{equation}
Here, $T$ (is equal to $\displaystyle{\tau^{-1}}$) stands for the ensemble temperature, and $\tau$ depicts the time period of the Euclidean. The generalized free energy achieves an on-shell condition solely when the time period of the Euclidean ($\tau$) matches the expression $\tau_{\mathrm{e}} = \frac{1}{T_{\mathrm{e}}}$. By following the procedure in Ref.~\cite{19a} with Eq.~(\ref{F1}), a vector field $\phi$ can be derived as 
\begin{equation}\label{F2}
\begin{split}
\phi=\left(\frac{\partial\mathcal{F}}{\partial r_{\mathrm{e}}},-\csc\Theta\cot\Theta\right)\,,
\end{split}
\end{equation}
where $\phi^{\Theta}$ diverges\footnote{One can also use the term $\frac{1}{\cos \Theta}$ instead  of $\frac{1}{\sin \Theta}$ but then the range for $\Theta$ will also modified to $-\pi/2\leq~\Theta\leq~\pi/2$ as described in \cite{Bhattacharya:2024bjp}.}, the orientation is outwards of the vector points at $\Theta = [\Theta,\pi]$. The ranges for $r_{\mathrm{e}}$ is $[\Theta,~\infty]$ and  for $\Theta$ is $[\Theta,~\pi]$. The topological current is defined by adopting the concept of Duan's $\phi$-mapping topological current theory, which is expressed as
\begin{equation}\label{F3}
j^{\tilde{\mu}}=\partial_{\tilde{\mu}}{V^{\tilde{\mu}\tilde{\nu}}}=\frac{1}{2\pi}\varepsilon^{\tilde{\mu}\tilde{\nu}\tilde{\rho}}\varepsilon_{ab}\partial_{\tilde{\nu}}n^{c}\partial_{\tilde{\rho}}n^{d}\,,
\end{equation}
where $\tilde{\mu},\tilde{\rho},\tilde{\nu}=0,1,2$. Here, $V_{\tilde{\mu}\tilde{\nu}}=\frac{1}{2\pi}\varepsilon^{\tilde{\mu}\tilde{\nu}\tilde{\rho}}\varepsilon_{cd}n^{c}\partial_{\tilde{\rho}}n^{d}$ is an anti-symmetric superpotential and $n^{c}=({n^{1},~n^{2}})=(\frac{\phi^r}{||\phi||},~\frac{\phi^\Theta}{||\phi||})$ is the normalized vector. \\
\indent Now, by utilizing Noether's theorem, the topological current that emerges is conserved, as follows
\begin{equation}\label{F4}
\partial_{\tilde{\mu}}j^{\tilde{\mu}}=0\,.
\end{equation}
We modified the topological current by making some adjustments to obtain the topological number
\begin{equation}\label{F5}
\begin{split}
j^{\tilde{\mu}}=\delta^{2}(\phi) J^{\tilde{\mu}}\left(\frac{\phi}{\mathbb{X}}\right)\,.
\end{split}
\end{equation}
The Jacobi tensor is characterized by the given expression
\begin{equation}\label{F6}
\begin{split}
\varepsilon^{cd}J^{\tilde{\mu}}\left(\frac{\phi}{\mathbb{X}}\right)=\varepsilon^{\tilde{\mu}\tilde{\nu}\tilde{\rho}}\partial_{\tilde{\nu}}\phi^{c}\partial_{\tilde{\rho}}\phi^{d}\,.
\end{split}
\end{equation}
The Jacobi vector simplifies to the classical Jacobi form when $\tilde{\mu}$ becomes zero, which can be seen from the expression $J^{0}\left(\frac{\phi}{\mathbb{X}}\right)=\frac{\partial \phi^{1}}{\partial {\mathbb{X}^{1}}},\frac{\partial \phi^{2}}{\partial {\mathbb{X}^{2}}}$. The equation indicates that $j^{\tilde{\mu}}$ becomes non-zero exclusively when $\phi$ equals zero. After doing some manipulations, we can represent the total charge $W$ as follows
\begin{equation}\label{F7}
\begin{split}
W=\int_{\Sigma}j^{0}d^2 \mathbb{X}=\sum_{i=1}^{n}\beta_{i}\eta_{i}=\sum_{i=1}^{n}\omega_{i}\,.
\end{split}
\end{equation}
Here, $\beta_i$ stands for the positive Hopf index, which quantifies which quantifies the loops formed by the vector $\phi^a$ within the $\phi$ space when $\mathbb{X}^{\tilde{\mu}}$ is close to the ZP $(z_{i})$. Moreover, $\eta_i$ identifies the direction or orientation corresponding to the topological framework at the ZP $(z_{i})$, which can be either positive or negative, as indicated by $\pm 1$. For the $i$-th ZP of $\phi$ in the region $\Sigma$, the associated winding number is given by $\omega_i$. It is important to recognize that $\omega_i$ remains unchanged regardless of the geometry of the region in which it is calculated. The winding number is closely connected to the stability of a BH, where a positive winding number indicates a stable state, while a negative value signifies an unstable BH configuration. It is quite easy to compute the generalized Helmholtz free energy for the BW BH by utilizing Eq.~(\ref{F1}), which takes the following shape
\begin{eqnarray}\nonumber
\mathcal{F}&=&\frac{1}{2 \pi ^{3/2} (-\delta -2)}\bigg\{S^{\frac{1}{\delta +2}} \bigg[\pi  \left(\beta ^2-1\right) (\delta +1)+\alpha ^2 \\\label{F8}&\times&(\delta -1) S^{\frac{2}{\delta +2}}+2 \sqrt{\pi } \alpha \ \beta \ \delta \ S^{\frac{1}{\delta +2}}\bigg]\bigg\}\,.
\end{eqnarray}

Furthermore, we mentioned here that to retrieve the generalized Helmholtz free energy for the Bekenstein-Hawking entropy; one can put the deformation parameter $\Delta=0$. We compute the thermodynamic potential $(\phi^{r_{\mathrm{e}}},~\phi^{\Theta})$ to discuss the thermodynamic topology in terms of \bre~by employing Eq.~(\ref{F2}) which takes the given shape
\begin{eqnarray}\nonumber
&\phi ^{r_\mathrm{e}}&=-3 \alpha ^2 \tau\bigg[\pi ^{\frac{\delta }{2}+1} \left(r_\mathrm{e}^{2}\right)^{\frac{\delta }{2}+1}\bigg]^{\frac{3}{\delta +2}}\hspace{-5 mm}-4 \sqrt{\pi } \alpha  \beta  \tau  \bigg[\pi ^{\frac{\delta }{2}+1} \hspace{-5 mm}\\\nonumber&\times&\hspace{-1 mm}\left(r_\mathrm{e}^{2}\right)^{\frac{\delta }{2}+1}\bigg]^{\frac{2}{\delta +2}}\hspace{-4 mm}-\pi  \left(\beta ^2-1\right) \tau  \left[\pi ^{\frac{\delta }{2}+1} \left(r_\mathrm{e}^{2}\right)^{\frac{\delta }{2}+1}\right]^{\frac{1}{\delta +2}}\hspace{-5 mm}\\\label{F99}&-&2 \pi ^{\frac{\delta +5}{2}} (\delta +2) \left(r_\mathrm{e}^{2}\right)^{\frac{\delta }{2}+1}\bigg/(2 \pi ^{3/2} r_\mathrm{e} \ \tau)\,,\\\label{F9}
&\phi ^{\Theta }&=-\frac{\cot \Theta}{\sin \Theta}\,.    
\end{eqnarray}

Moreover, by using the above equation, we obtained the unit vectors $n^{1},~n^{2}$. To proceed, we determine the ZPs of the component $\phi^{r_\mathrm{e}}$ by employing the expression $\phi^{r_\mathrm{e}} = 0$ and subsequently compute the relation for $\tau$ as shown below 
\begin{eqnarray}\nonumber
\tau &=&\Bigg\{-2 \pi ^{\frac{\delta }{2}+\frac{5}{2}} (\delta +2) \left(r_\mathrm{e}^{2}\right)^{\frac{\delta }{2}+1}\Bigg\}\Bigg\{\bigg[\pi ^{\frac{\delta }{2}+1}\left(r_\mathrm{e}^{2}\right)^{\frac{\delta }{2}+1} \\\nonumber&\times&3 \ \alpha ^2 \bigg]^{\frac{3}{\delta +2}}+4 \ \sqrt{\pi } \ \alpha \ \beta \  \bigg[\pi ^{\frac{\delta }{2}+1} \left(r_\mathrm{e}^{2}\right)^{\frac{\delta }{2}+1}\bigg]^{\frac{2}{2+\delta}}\\\nonumber&+&\pi  \beta ^2 \bigg[\pi ^{\frac{\delta }{2}+1} \left(r_\mathrm{e}^{2}\right)^{\frac{\delta }{2}+1}\bigg]^{\frac{1}{\delta +2}}-\bigg[\pi ^{\frac{\delta }{2}+1} \left(r_\mathrm{e}^{2}\right)^{\frac{\delta }{2}+1}\bigg]^{\frac{1}{2+\delta}}\hspace{-5 mm}\\\label{F10}&\times&\pi\Bigg\}^{-1}.
\end{eqnarray}

Here, we delve into the valuable points related to thermodynamic topology for BW BHs. Among the freedom parameters discussed in this article, the dark matter parameter $\beta$ plays the most critical role in determining the number of topological charges. In our analysis, we utilize the cosmological constant $\alpha=0.7$ and the dark matter parameter $\beta=-1.4$ consistent with the BW BH framework discussed in Ref.~\cite{Shahzad:2021nqt}. These parameters represent the impact of effective modifications caused by the bulk on the four-dimensional geometry. To maintain thermodynamic consistency and facilitate a comprehensive study of BH stability, the chosen values are physically well-founded, ensuring the presence of both event and cosmological horizons. Furthermore, these values fall within the phenomenologically acceptable range discussed in earlier studies on bulk-generated dark energy and BW corrections \cite{Heydar-Fard:2007ahl,Maartens:2010ar,Shahzad:2021nqt}.

\begin{figure*}
     \centering
     \begin{subfigure}[b]{0.4\textwidth}
         \centering
         \includegraphics[height=5cm,width=7cm]{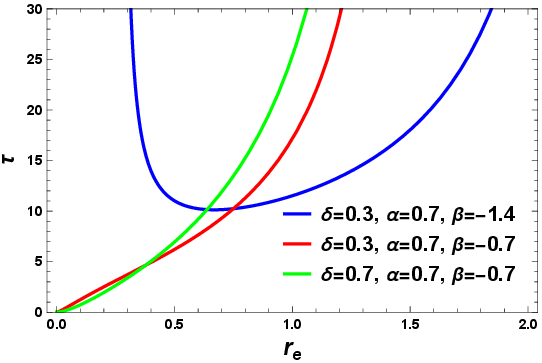}
         \caption{}\label{1a}
     \end{subfigure}
     \begin{subfigure}[b]{0.4\textwidth}
         \centering
         \includegraphics[height=5cm,width=7cm]{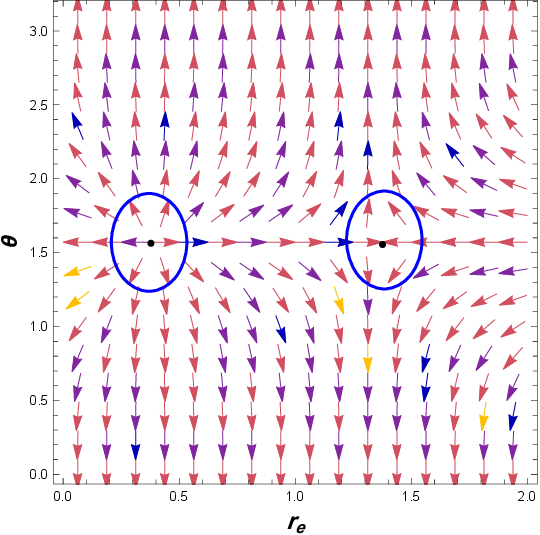}
         \caption{}\label{1b}
     \end{subfigure}
      \begin{subfigure}[b]{0.4\textwidth}
         \centering
         \includegraphics[height=5cm,width=7cm]{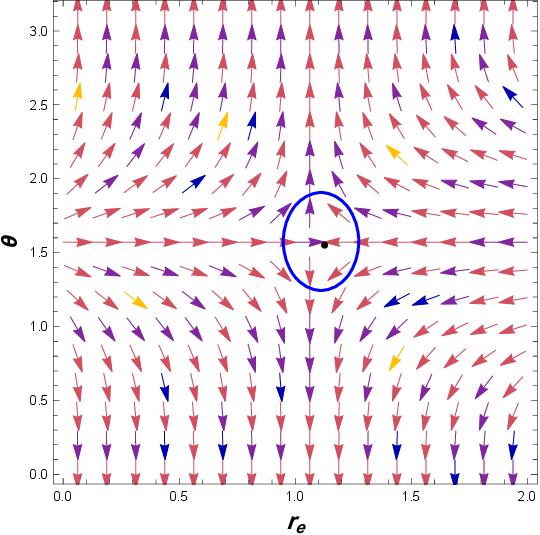}
         \caption{}\label{1c}
     \end{subfigure}
      \begin{subfigure}[b]{0.4\textwidth}
         \centering
         \includegraphics[height=5cm,width=7cm]{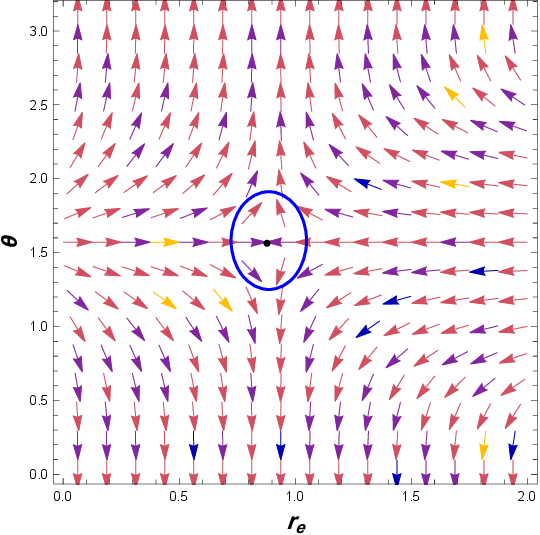}
         \caption{}\label{1d}
     \end{subfigure}\\
        \caption{\raggedright In Fig.~\ref{m1}, rainbow color  arrows presented  the normalized vector in $r_{\mathrm{e}}-\Theta$ plane for \bre. The trajectory defined by Eq.~(\ref{F10}) is showed in Fig.~\ref{m1}(\subref{1a}). In Figs.~\ref{m1}(\subref{1b}),~\ref{m1}(\subref{1c}),~\ref{m1}(\subref{1d}), the ZPs or (phase transition) are demonstrated at coordinates $(r_e, \Theta)$ along the circular paths which are determined by the free parameter associated with \bre.}
        \label{m1}
\end{figure*}

\begin{figure*}
     \centering
      \begin{subfigure}[b]{0.4\textwidth}
         \centering
         \includegraphics[height=5cm,width=7cm]{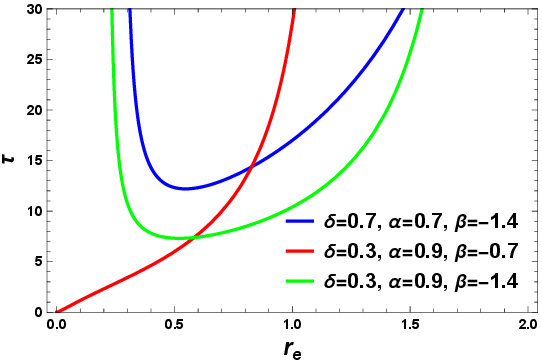}
         \caption{}\label{1g}
     \end{subfigure}
      \begin{subfigure}[b]{0.4\textwidth}
         \centering
         \includegraphics[height=5cm,width=7cm]{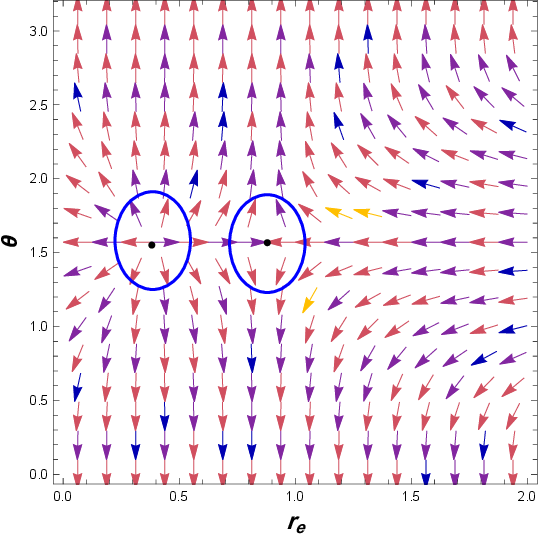}
         \caption{}\label{1h}
     \end{subfigure}
      \begin{subfigure}[b]{0.4\textwidth}
         \centering
         \includegraphics[height=5cm,width=7cm]{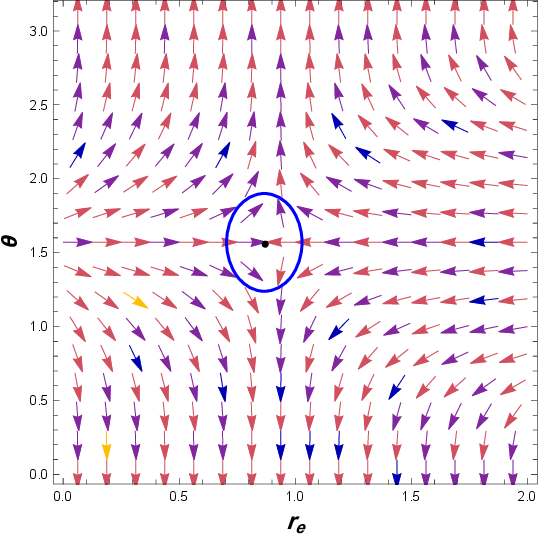}
        \caption{} \label{1i}
     \end{subfigure}
      \begin{subfigure}[b]{0.4\textwidth}
         \centering
         \includegraphics[height=5cm,width=7cm]{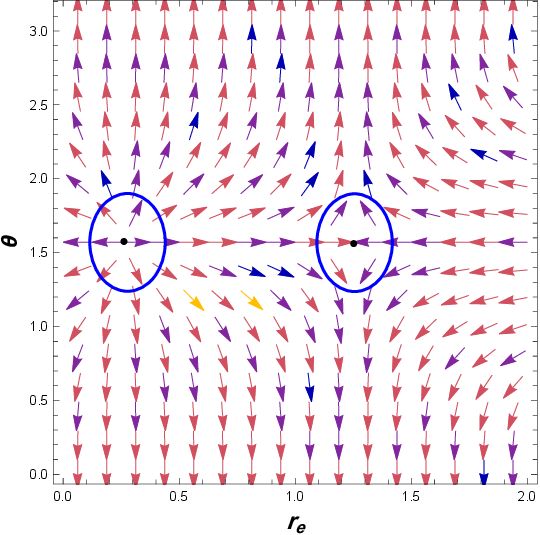}
        \caption{} \label{1j}
     \end{subfigure}
        \caption{\raggedright In Fig.~\ref{m11}, rainbow color  arrows presented  the normalized vector in $r_{\mathrm{e}}-\Theta$ plane for \bre. A trajectory defined by Eq.~(\ref{F10}) is showed in Fig.~\ref{m11}(\subref{1g}). While, in Figs.~\ref{m11}(\subref{1h}),~\ref{m11}(\subref{1i}) and \ref{m11}(\subref{1j}), the ZPs or (phase transition) are demonstrated at coordinates $(r_e, \Theta)$ along the close paths which are computed by the free parameter associated with \bre.}
        \label{m11}
\end{figure*}

\begin{figure*}
     \centering
     \begin{subfigure}[b]{0.3\textwidth}
         \centering
         \includegraphics[height=5cm,width=5cm]{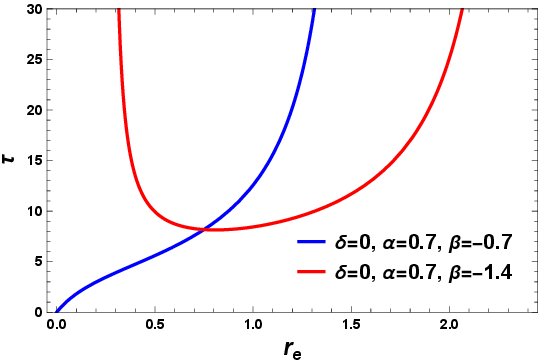}
         \caption{}
         \label{2a}
     \end{subfigure}
     \begin{subfigure}[b]{0.3\textwidth}
         \centering
         \includegraphics[height=5cm,width=5cm]{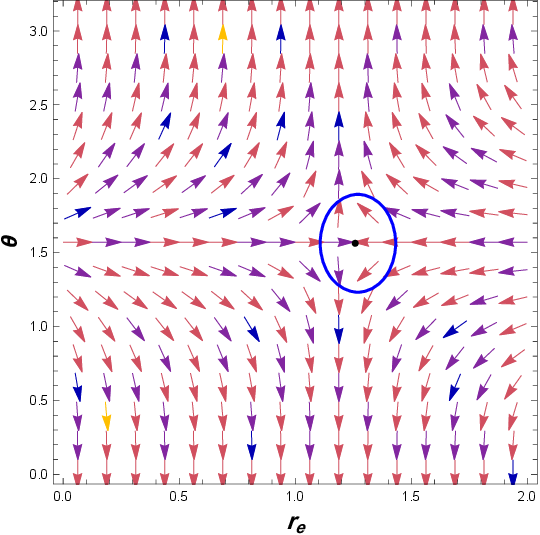}
         \caption{}
         \label{2b}
     \end{subfigure}
      \begin{subfigure}[b]{0.3\textwidth}
         \centering
         \includegraphics[height=5cm,width=5cm]{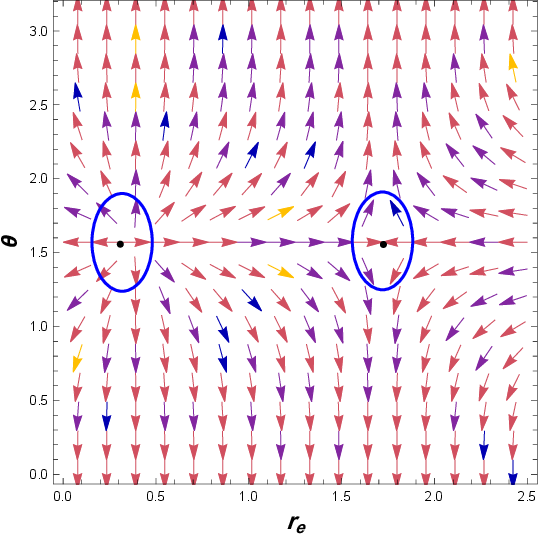}
         \caption{}
         \label{2c}
     \end{subfigure}
        \caption{\raggedright In Fig.~\ref{m2}, rainbow color  arrows depicted the normalized vector in $r_{\mathrm{e}}-\Theta$ plane for the Hawking-Bekenstein entropy. The curve defined by Eq.~(\ref{F10}) is depicted in Fig.~\ref{m2}(\subref{2a}). In Figs.~\ref{m2}(\subref{2b}) and \ref{m2}(\subref{2c}), the ZPs (phase transition) are displayed at coordinates $(r_e, \Theta)$ along the circular paths, which are determined by the free parameter when $\delta=0$.}
        \label{m2}
\end{figure*}

For instance, as illustrated in Fig.~\ref{m1}(\subref{1b}), when we set the free parameters to ($\delta = 0.3$), ($\alpha = 0.7$), and ($\beta = -1.4$), we find two topological charges $(\omega=~+1,~-1)$ with a total topological charge of ($W = 0$)\footnote{ One can obtain the total topological charge by adding the winding number which is corresponding to the number of phase transition or ZPs \cite{50a}. For example, in our case, we have obtained two ZPs or phase transitions corresponding to which we have winding numbers -1 and 1. Thereby, by adding these winding numbers, we can obtain a total topological charge of $W=0$.}. However, by reducing the parameter ($\beta$), the number of topological charges also decreases. This is evident in Fig.~\ref{m1}(\subref{1d}), where, with ($\delta$) and ($\alpha$) held constant and ($\beta$) reduced, the number of topological charges decreases to one $(\omega=~-1)$, resulting in a total topological charge of ($W=~-1$). Interestingly, increasing the parameter ($\delta$), as shown in Fig.~\ref{m1}(\subref{1b}), or altering the parameter ($\alpha$), as depicted in Figs.~\ref{m11}(\subref{1h}) and \ref{m11}(\subref{1j}), do not affect the number of topological charges. This shows the role of the parameter ($\beta$) in influencing the topological charges, as further illustrated in Figs.~\ref{m1} and \ref{m11}. These findings are summarized in Table~\ref{P1}.
A similar result is observed when the deformation parameter ($\delta$=0), as shown in Figs.~\ref{m2}(\subref{2b}) and \ref{m2}(\subref{2c}), highlighting the significant impact of the parameter ($\beta$) on the number of topological charges in the current structure. In Figs.~\ref{m1}(\subref{1a}),~\ref{m11}(\subref{1a}),~\ref{m2}(\subref{2a}), we have plotted the trajectory corresponding to Eq.~(\ref{F10}) across various values of the free parameters. When mapping the free energy function as a scalar quantity in the two-dimensional space \((r_e, \Theta)\), the resulting vector field \((\phi)\) is formulated in such a way that the extremum points of the free energy function act as ZPs for this vector field. In this context, the behavior of the field lines of \((\phi)\) around these ZPs, depending on whether they correspond to a maximum or minimum in the free energy function, exhibits rotational characteristics. The structure of these rotations allows for the assignment of a topological charge to each of these ZPs \cite{19a}.
Preliminary studies on fundamental BHs, such as Schwarzschild and Reissner-Nordstr\"{o}m BHs, have demonstrated that the total topological charge for these BHs follows a distinct pattern: For a Schwarzschild BH, the total topological charge is \(W = -1\). For a Reissner-Nordström BH, the total topological charge is \(W = 0\). For an AdS-Reissner-Nordstr\"{o}m BH, the total topological charge is \(W = +1\) \cite{19a}.
Since these BHs serve as classical foundations in BH physics, their behavior is utilized as a general model for categorizing BH dynamics, including thermodynamics. As shown in Table (\ref{S1}), BHs with a total topological charge of \(-1\) exhibit Schwarzschild-like behavior, whereas those classified with \(W = 0\) belong to the Reissner-Nordstr\"{o}m category. Accordingly, our calculations yield a range of results that, while diverse across different parameter intervals, confirm that within various free parameter domains in the BW scenario, BH behavior aligns well with the established topological charge classifications. Specifically, in our model, \(W = -1\) corresponds to Schwarzschild-like behavior, whereas \(W = 0\) matches the characteristics of a Reissner-Nordstr\"{o}m BH, as demonstrated in Table (\ref{P1}) and in these two cases, considering the shape of the $\tau$ diagram, it seems unlikely that a first-or second-order Van der Waals phase transition will occur in the system. \\
\indent Also, the dark matter parameter $\beta$ is a critical variable in this model, often representing the contribution of dark matter to the BH's overall structure. The $\beta$-parameter can play an essential role in shaping the spacetime geometry and influencing the distribution of topological charges. In these calculations with respect to the Table (\ref{P1}), a change in the topological charge results from decreasing and increasing this parameter; it can be shown that changes in this parameter are effective in the number of topological charges, and this view is shown by observing the changes in topological charges in the figures. The claim suggests that the dark matter component (or its parameter $\beta$) is also important for determining how many distinct topological charges the BH can exhibit. This means that as $\beta$ changes, it alters the topological charges of the BH, which in turn affects the winding number and the total number of topological charges. If $\beta$ increases or decreases, it could shift the BH between different thermodynamic phases, each of which corresponds to different topological charges. Additionally, for a more effective comparison of our results with those obtained in other studies, please refer to Tables (\ref{P1}) and (\ref{S1}).

\begin{table}[t]
\caption{\label{P1}\raggedright Summary of the topological classification of BW BHs on the basis of topological charge (or winding number) for numerous values of $\alpha,~\delta$, and $\beta$.}
\begin{ruledtabular}
\begin{tabular}{ccccc}
   $\delta$ & $\alpha$ &  $\beta$ & $\omega$ & $W$ \\
   \hline
    0.3 & 0.7& -1.4 &  +1, -1 & 0 \\
    0.3&  0.7&  -0.7 & -1 &   -1  \\
    0.7&  0.7&  -0.7 &    -1 &   -1 \\
    0.7&  0.7&  -1.4 &   +1, -1&   0 \\
     0.3&  0.9&  -0.7 &    -1 &  -1 \\
    0.3&  0.9&  -1.4 &  +1, -1 &    0 \\
   0&  0.7&  -0.7 &  -1 &  -1 \\
   0 &  0.7 &   -1.4 &    +1, -1 &   0  \\
  \hline
\end{tabular}
\end{ruledtabular}
\end{table}
\begin{table}[h]
    \centering
    \caption{\label{S1}\raggedright Topological Number $W$ for Various BHs}
    \begin{tabular}{|l|c|}
        \hline
        \textbf{Black Hole Solution} & \textbf{Topological Charge $W$}\\
        \hline
        Schwarzschild-AdS$_4$ BH \cite{21a} & 0 \\
        Schwarzschild BH \cite{19a} & -1 \\
        Reissner-Nordström BH \cite{19a} & 0 \\
        Reissner-Nordström AdS BH & 1 \\
        Kerr BH \cite{Wu:2022whe,41a} & 0 \\
        Kerr-Newman BH \cite{Wu:2022whe,41a} & 0 \\
        $d = 5$ Singly Rotating Kerr BH \cite{Wu:2022whe,41a} & 0 \\
        $d \geq 6$ Singly Rotating Kerr BH \cite{Wu:2022whe,41a} & -1 \\
        \hline
    \end{tabular}
\end{table}

In black hole thermodynamics, the free energy can be expressed as a scalar function on a two-dimensional parameter space, typically spanned by the horizon radius \( r_{e} \) and a complementary coordinate \( \Theta \). From this function, a vector field \( \boldsymbol{\phi} \) is derived as its gradient. Equilibrium states of the system correspond to the zeros of \( \boldsymbol{\phi} \), where the gradient vanishes.

The local behavior of \( \boldsymbol{\phi} \) around these equilibrium points carries topological information. Each zero is assigned a topological index—or winding number—which quantifies the circulation of the vector field around it. This integer index reflects the nature of the extremum in the free energy: positive winding generally indicates a stable configuration, whereas negative winding signals instability.

This topological classification serves as a universal framework for characterizing black hole solutions. For instance, the Schwarzschild black hole possesses a total winding number \( W = -1 \), the Reissner–Nordstr\"{o}m solution yields \( W = 0 \), and the AdS–Reissner–Nordstr\"{o}m black hole exhibits \( W = +1 \). These distinct values correspond to fundamentally different thermodynamic behaviors and phase structures. Solutions with \( W = +1 \) typically display AdS-like stability properties, while those with \( W = 0 \) align with Reissner–Nordström-type phase behavior.

As parameters such as charge, $\beta$, or $\delta$ parameters are varied, the distribution of zeros in \( \boldsymbol{\phi} \) evolves. Pairs of zeros—one stable and one unstable—can be created or annihilated, but the total winding number remains conserved. This conservation reflects a topological constraint on possible phase transitions. The sensitivity of the winding number to parameter changes, such as those induced by quantum corrections or higher-dimensional modifications, reveals how gravitational extensions reshape the thermodynamic landscape.

By evaluating the winding number across parameter space, one can systematically predict the emergence, stability, and classification of black hole phases. The approach not only distinguishes between first- and second-order phase transitions but also generalizes to broader classes of black holes in modified gravity. Ultimately, the topological winding number acts as a robust invariant that encodes the global structure of thermodynamic phase space, offering a powerful diagnostic for understanding black hole thermodynamics in extended theories of gravity. We indeed established that all three primary parameters—the cosmological parameter \(\alpha\), the quantum-gravity deformation parameter \(\delta\), and the dark matter parameter \(\beta\)—play crucial and non-trivial roles in shaping the local thermodynamic behavior of the BW black hole. As detailed in our study of heat capacity and thermodynamic geometry, each parameter significantly affects the location of stability bounds (roots) and phase transition points (divergences). The Ricci scalar of the Ruppeiner metric, a sensitive probe of microscopic interactions and phase structure, confirmed that singularities shift systematically with variations in \(\alpha\), \(\delta\), and \(\beta\). This demonstrates that the local thermodynamic geometry is a collective function of the entire parameter set.
However, when progressing to the global, topological analysis—which classifies the entire thermodynamic phase space via the topological charge \(W\)—a distinct hierarchy of influence emerges. Our systematic investigation reveals the following critical pattern: while varying \(\alpha\) or \(\delta\) with \(\beta\) fixed can alter the characteristics of critical points (e.g., their horizon radius or temperature), these variations do not typically induce a change in the integer value of the total topological charge \(W\). The phase space structure, in terms of the number and fundamental type of stable and unstable phases, remains topologically invariant under such changes.
Conversely, varying the dark matter parameter \(\beta\) proves to be the most potent trigger for a global topological phase transition. For fixed values of \(\alpha\) and \(\delta\), adjusting \(\beta\) beyond critical thresholds reliably switches the total charge between values such as \(W = -1\) and \(W = 0\). This signifies a fundamental restructuring of the free-energy landscape, where entire solution branches are created or annihilated. The ``dominance" of \(\beta\) is therefore defined in this specific, rigorous sense: it is the parameter whose variation most directly and consistently governs changes in the global topological invariant of the system, a property more robust and profound than shifting local critical points.
This exceptional role of \(\beta\) is not a contradiction but a deeper insight stemming from the mathematical structure of the braneworld metric given in Eq.~\eqref{METF}. The parameter \(\beta\) uniquely appears in a constant, additive term \((-\beta^2)\). This term acts as a universal shift in the gravitational potential, effectively altering the asymptotic boundary condition for the free energy construction used in the topological method. Parameters like \(\alpha\) and \(\delta\), while critically important, often enter in a more functional, scale-dependent manner (e.g., \(\alpha^2 r^2\)). Their variations modify the ``shape" of the thermodynamic landscape, but the constant shift from \(\beta\) can change its very ``foundation," thereby controlling the topological class. Physically, this underscores that \(\beta\), encoding the tidal charge from the extra dimension, impacts the gravitational binding in a more global, integrated manner compared to the more localized or environmental effects of \(\alpha\) (cosmological background) and \(\delta\) (horizon microstructure).
Thereby, our findings present a coherent two-tiered picture:
1.  Thermodynamics (Heat Capacity, Ruppeiner Geometry): Governed by a strong, interconnected interplay of all parameters \(\alpha\), \(\delta\), and \(\beta\).\\
2. Thermodynamic Topology (Topological Charge \(W\)): Primarily and most sensitively governed by the parameter \(\beta\), which holds the key to transitions between distinct topological universality classes.
\section{Topological Photon spheres}\label{Sec-6}

Before delving into the topological structure of the photon sphere, it is prudent first to examine the structure of the length element and the metric function. Clearly, the four-dimensional length element Eq.~(\ref{LINE}) of this structure exhibits a static form and spherical symmetry \cite{Heydar-Fard:2007ahl}. However, it is noteworthy that the metric function Eq.~(\ref{METF}) in its asymptotic form (with non-zero $\alpha$ and $\beta$) can only follow a dS structure. Since the radius tends to infinity, the dominant term in the metric function will be $-r^2$. In other words, to achieve an AdS form, $\alpha$ must be considered imaginary, which seems to lack physical meaning. Also, an asymptotically flat state would result in zeroing the parameters, which, as previously discussed, effectively transforms the model into the Schwarzschild state.\\
\indent Unlike models previously examined using topological methods \cite{33a,60a,61a, 610a}, which were either asymptotically flat or in AdS form, it is important to note that in the dS structure, a cosmological horizon ($r_\mathrm{c}$) always appears. Due to its relatively large radius compared to the event horizon, this horizon effectively inhibits the creation of a photon sphere beyond it. In other words, the effective potential beyond the cosmological horizon lacks real value. This was first mentioned in Wei's paper in the form of a topological study of the photon sphere \cite{50a}. In this regard, subsequent studies have also confirmed such behavior \cite{3100,3200,3300}. Therefore, for studying such structures, we consider the space $r_\mathrm{h} \leq r \leq r_\mathrm{c}$. By adopting the methodology discussed in Refs.~\cite{33a, 50a, 60a, 61a}, we have computed the standard potential, which is given as
\begin{equation}\label{Ph1}
H(r,\theta)=\sqrt{\frac{-g_{\mathrm{t}\mathrm{t}}}{g_{\varphi\varphi}}}=\frac{1}{\sin\theta}\left(\frac{f(r)}{h(r)}\right)^{1/2}\,,
\end{equation}
where $H(r,\theta)$ is the effective potential for photon orbits, $f(r)$ and $h(r)$ are the metric functions of the time component and angular component in spherical coordinates. Here, we mention that in the case of a spherically symmetric metric and with respect to Eq. (\ref{METF}), we put $h(r)=~r^{2}$ and $-g_{tt}=f(r)=A(r)$. By analyzing this potential, we can find the radius of the photon sphere, which is determined by $\partial_{r} H=0$. Furthermore, it is easy to define vector field $(\phi^r, \phi^\theta)$ by  employing Eq.~(\ref{Ph1}) which is written as
\begin{equation}\label{Ph2}
\phi^r=\frac{\partial_r H}{\sqrt{g_{rr}}}=\sqrt{g(r)}\partial_r H\,, \quad \phi^\theta=\frac{\partial_\theta H}{\sqrt{g_{\theta\theta}}}=\frac{\partial_\theta H}{\sqrt{h(r)}}\,.
\end{equation}
The total charge is then given by
\begin{equation}\label{Ph3}
Q=\sum_{i}\omega_i\,.
\end{equation}
Thus, it is observed that the ZP enclosed in a curve predicted that the charge $Q$ is equal to the winding number (for more details, see Ref.~\cite{61a}). Fortunately, the metric function can be solved according to the parameters and has the following general form as follows
\begin{eqnarray}\nonumber
&r_{\mathrm{h},\mathrm{c}}& =\varepsilon_{1}+\varepsilon_{2}\,,\\\nonumber
&\varepsilon_{1}&=\frac{1}{{3 \alpha}}\bigg[\beta^{3}+3 \sqrt{3}\, \bigg(-2 \alpha  \,\beta^{3} M +27 \alpha^{2} M^{2}-\beta^{4}\\\nonumber&+& \hspace{-3mm}18 \alpha  \beta  M +2 \beta^{2}-1\bigg)^{1/2}-27 M \alpha -9 \beta \bigg]^{\frac{1}{3}}\,,\\\nonumber
&\varepsilon_{2}&=\frac{\beta^{2}+3}{3 \alpha}\bigg\{ \bigg[\beta^{3}+3 \sqrt{3}\, \bigg(-2 \alpha  \,\beta^{3} M +27 \alpha^{2} M^{2}\\\nonumber&-&\hspace{-3mm}\beta^{4}+18 \alpha  \beta  M +2 \beta^{2}-1\bigg)^{\frac{1}{2}}\hspace{-2mm}-27 M \alpha-9 \beta \bigg]^{\frac{1}{3}}\bigg\}^{-1}\hspace{-3mm}\\\label{Ph4}&-&\frac{2 \beta}{3 \alpha}\,.
\end{eqnarray}
The generic form of the functions for the BW BH can be obtained by utilizing Eqs.~(\ref{Ph1}) and (\ref{Ph2}), which is given as 
\begin{eqnarray}\label{Ph5}
&H &=\frac{\sqrt{1-\frac{2 M}{r}-r^{2} \alpha^{2}-2 \beta  r \alpha -\beta^{2}}}{r \sin \! \theta }\,,\\\label{Ph6}
&\phi^{r}&=\frac{\left(\beta  r^{2} \alpha +\beta^{2} r +3 M -r \right) \csc \! \theta}{r^{3}}\,,\\\label{Ph7}
&\phi^{\theta}&=-\frac{\sqrt{1-\frac{2 M}{r}-r^{2} \alpha^{2}-2 \beta  r \alpha -\beta^{2}}\, \cos \! \theta }{(r \sin\! \theta)^{2}}\,.
\end{eqnarray}
\indent To better understand the impact of the parameters on the photon sphere, we study the BW  BH in three different scenarios. We have also constructed a table in  Table~\ref{P2} to analyze the ranges of parameters for the BW BH for these cases. It is worth mentioning here that in Table~\ref{P2}, FP is the fixed parameter, TTC is the total topological charge, and UA is an unauthorized region: the area where the roots computed from the $\phi$ equations turn out to be imaginary or negative, and UPS represents the unstable photon sphere.
\subsubsection{Case I: $\alpha\leq 0.05$}

If we consider $m=1$ and $\alpha = 0.05$, then the relation for the horizon radius in the form of dark matter parameter $\beta$ can be represented as
\begin{eqnarray}\nonumber
&r_{\mathrm{h},\mathrm{c}}&=\varepsilon_{1}+\varepsilon_{2}\,,\\\nonumber
&\varepsilon_{1}&= 6.666666667 \bigg[\beta^{3}+3 \sqrt{3}\, \bigg(-\beta^{4}- 0.10 \beta^{3}\\\nonumber&+&2 \beta^{2}+ 0.90 \beta - 0.9325\bigg)^{1/2}- 1.35-9 \beta \bigg]^{\frac{1}{3}}\,,\\\nonumber
&\varepsilon_{2}&=\bigg[6.666666667 \left(\beta^{2}+3\right)\bigg]\bigg\{\bigg[\beta^{3}+3 \sqrt{3}\, \\\nonumber&\times&\bigg(-\beta^{4}- 0.10 \beta^{3}+2 \beta^{2}+ 0.90 \beta - 0.9325\bigg)\\\label{Ph8}&-& 1.35-9 \beta \bigg]^{\frac{1}{3}}\bigg\}^{-1}- 13.33333333 \beta\,.  
\end{eqnarray}

If we draw $\beta$ in terms of $r$, we will see that the allowed range of $\beta$, Fig.~\ref{m3}(\subref{3a}) (we presented Fig.~\ref{m3} in the Appendix A), to have a BH form and satisfy the weak cosmic censorship conjecture (WCCC) will be $\beta\leq 5.2342\times10^{-1}$, which in Fig.~\ref{m3}(\subref{3b}) we have drawn some samples with the horizon and without horizon for the metric function by putting $\beta=0.001$ (green curve), $\beta=0.0025$ (light blue dashed curve), $\beta=0.05$ (dashed green curve), and $\beta=0.059$ (dotted dashed orange curve), and $\beta=0.09$ (dotted red curve). In this form, the smaller horizon is the event horizon ($r_{\mathrm{h}}$), and the larger horizon is the cosmological horizon ($r_{\mathrm{c}}$).

As a sample, for $\beta=0.04$, the general form of the main functions is

\begin{eqnarray}\label{Ph9}
&H& =\frac{\sqrt{ 0.9984-\frac{2}{r}- 0.0025 r^{2}- 0.0040 r}}{r}\,,\\\label{Ph10}&\phi^{r}&=\frac{\left(- 0.9984 r + 0.0020 r^{2}+3\right) \csc \! \theta }{r^{3}}\,,\\\nonumber
&\phi^{\theta}&=-\sqrt{ 0.9984-\frac{2}{r}- 0.0025 r^{2}- 0.0040 r}\\\label{Ph11}&\times&\frac{\cos \! \theta}{ \left(r \sin \!\theta\right)^{2}}\,.    
\end{eqnarray}

\begin{figure*}
     \centering
     \begin{subfigure}[b]{0.3\textwidth}
         \centering
         \includegraphics[width=\textwidth]{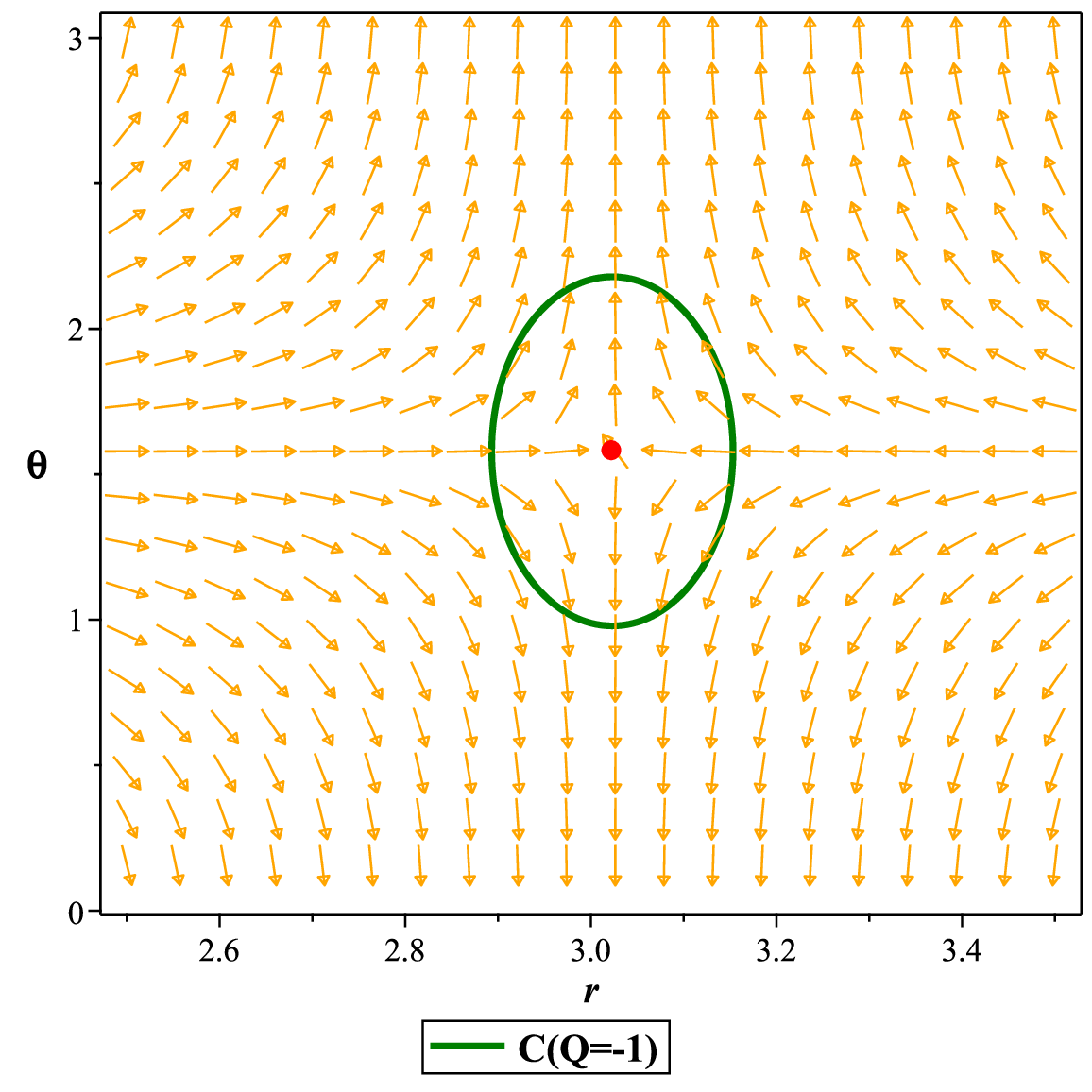}
         \caption{}
         \label{4a}
     \end{subfigure}
     \begin{subfigure}[b]{0.3\textwidth}
         \centering
         \includegraphics[width=\textwidth]{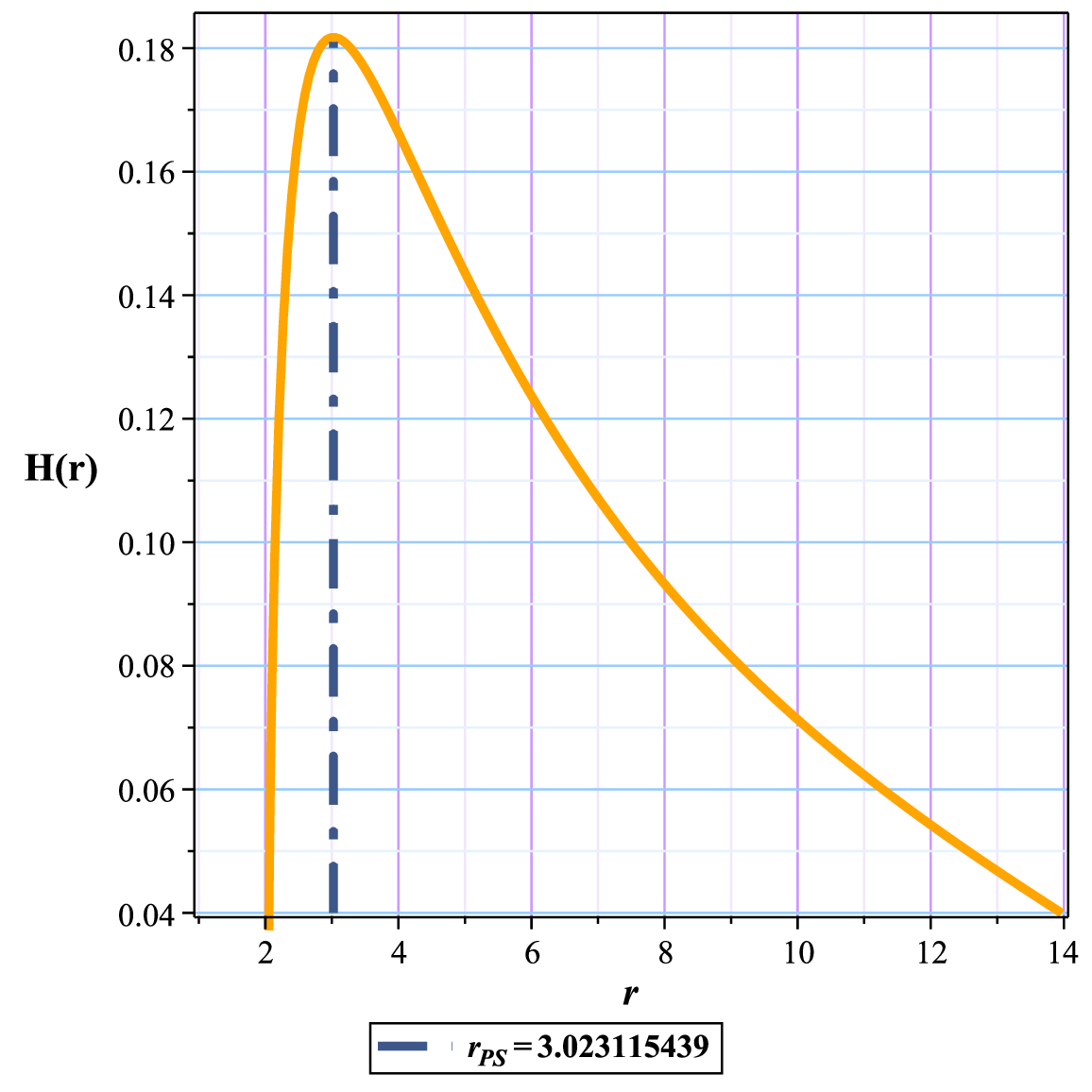}
         \caption{}
         \label{4b}
     \end{subfigure}
      \begin{subfigure}[b]{0.3\textwidth}
         \centering
         \includegraphics[width=\textwidth]{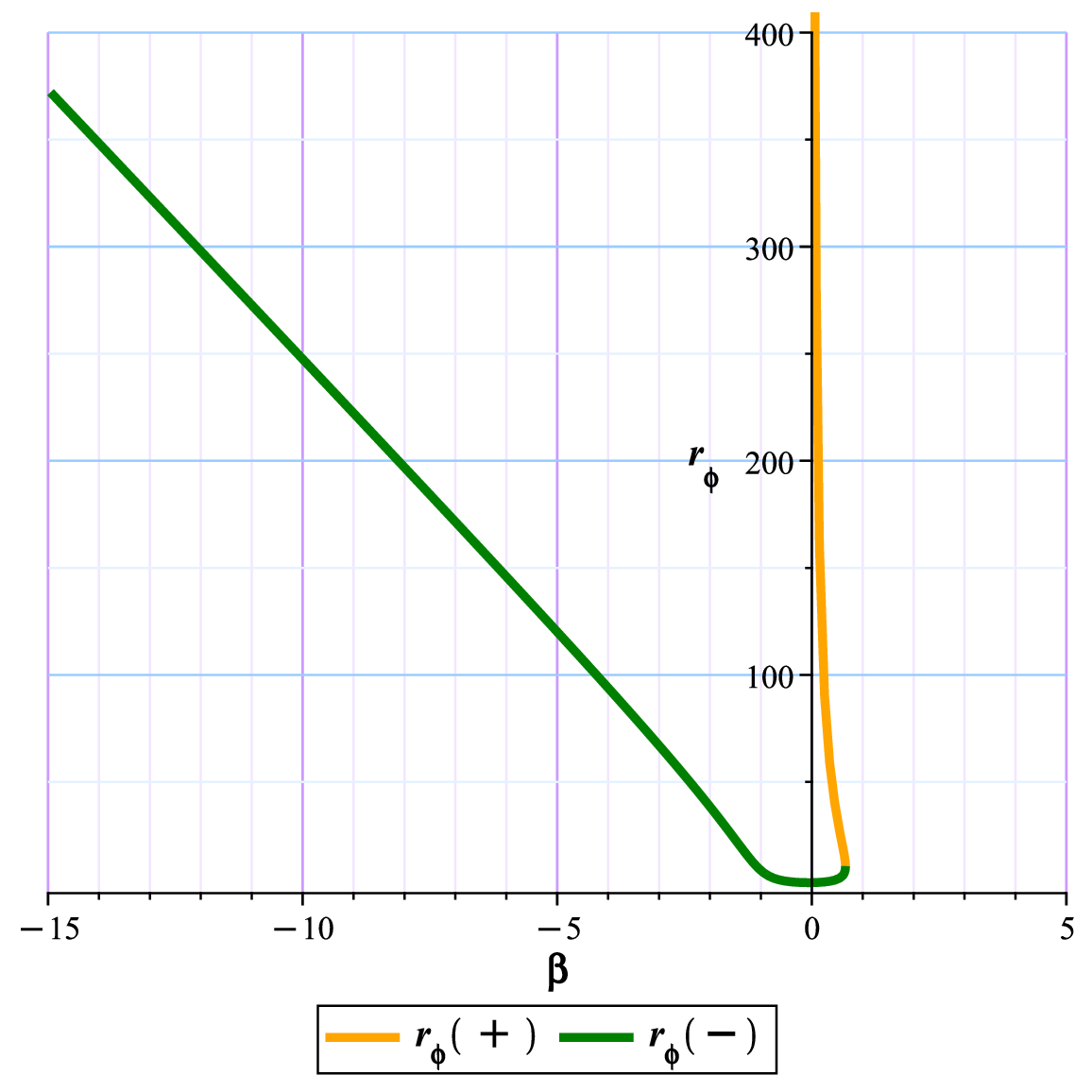}
         \caption{}
         \label{4c}
     \end{subfigure}
     \begin{subfigure}[b]{0.3\textwidth}
         \centering
         \includegraphics[width=\textwidth]{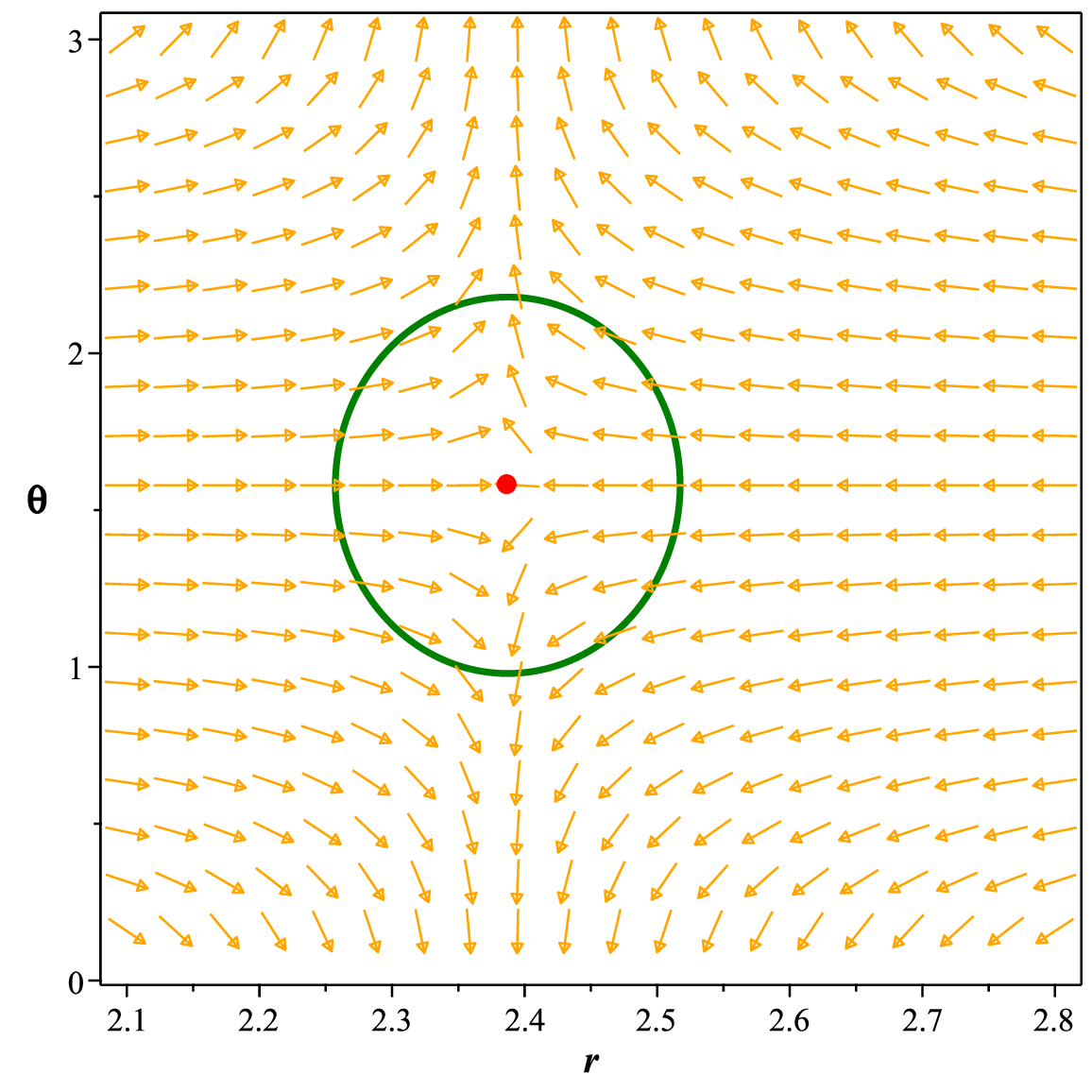}
         \caption{}
         \label{6a}
     \end{subfigure}
     \begin{subfigure}[b]{0.3\textwidth}
         \centering
         \includegraphics[width=\textwidth]{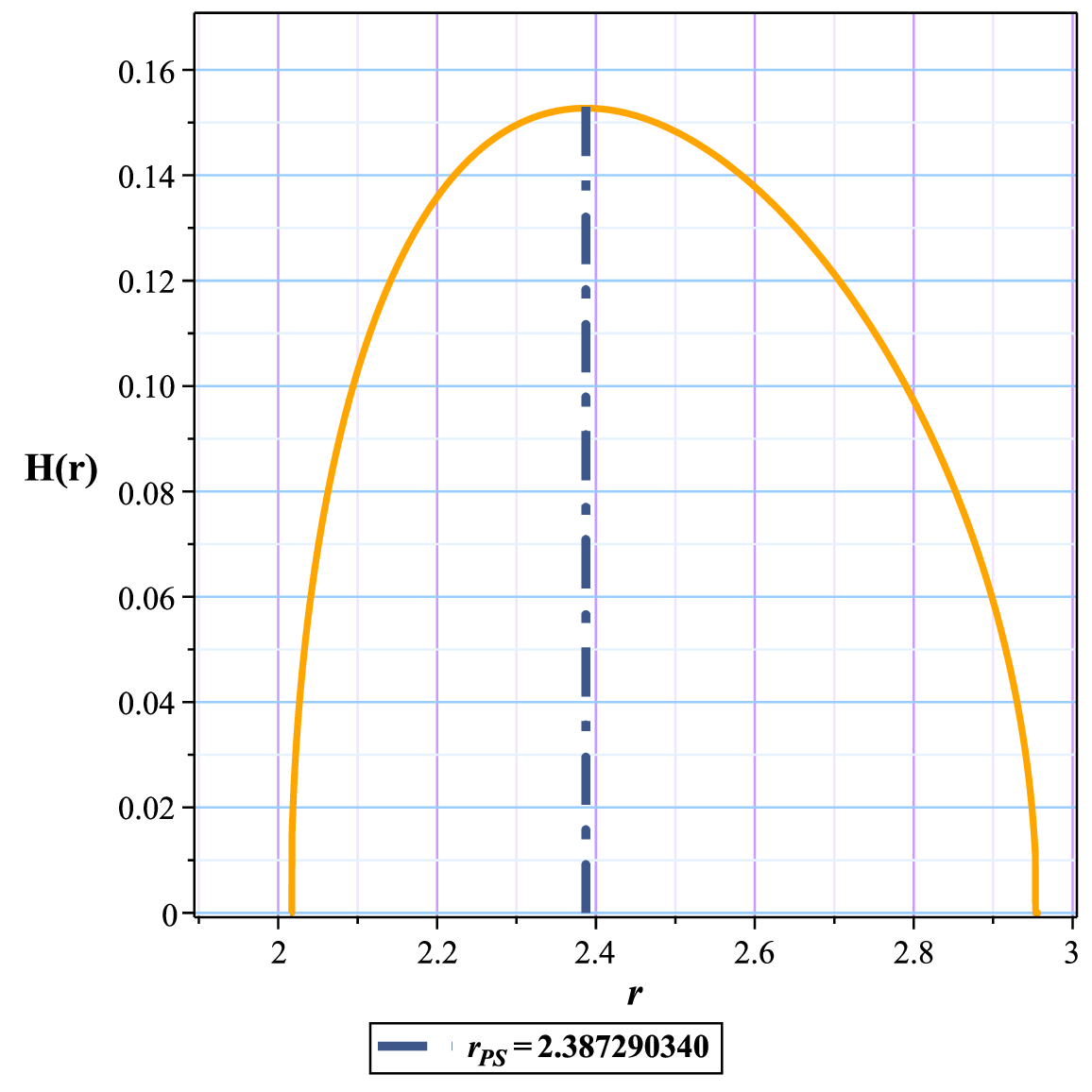}
         \caption{}
         \label{6b}
     \end{subfigure}
      \begin{subfigure}[b]{0.3\textwidth}
         \centering
         \includegraphics[width=\textwidth]{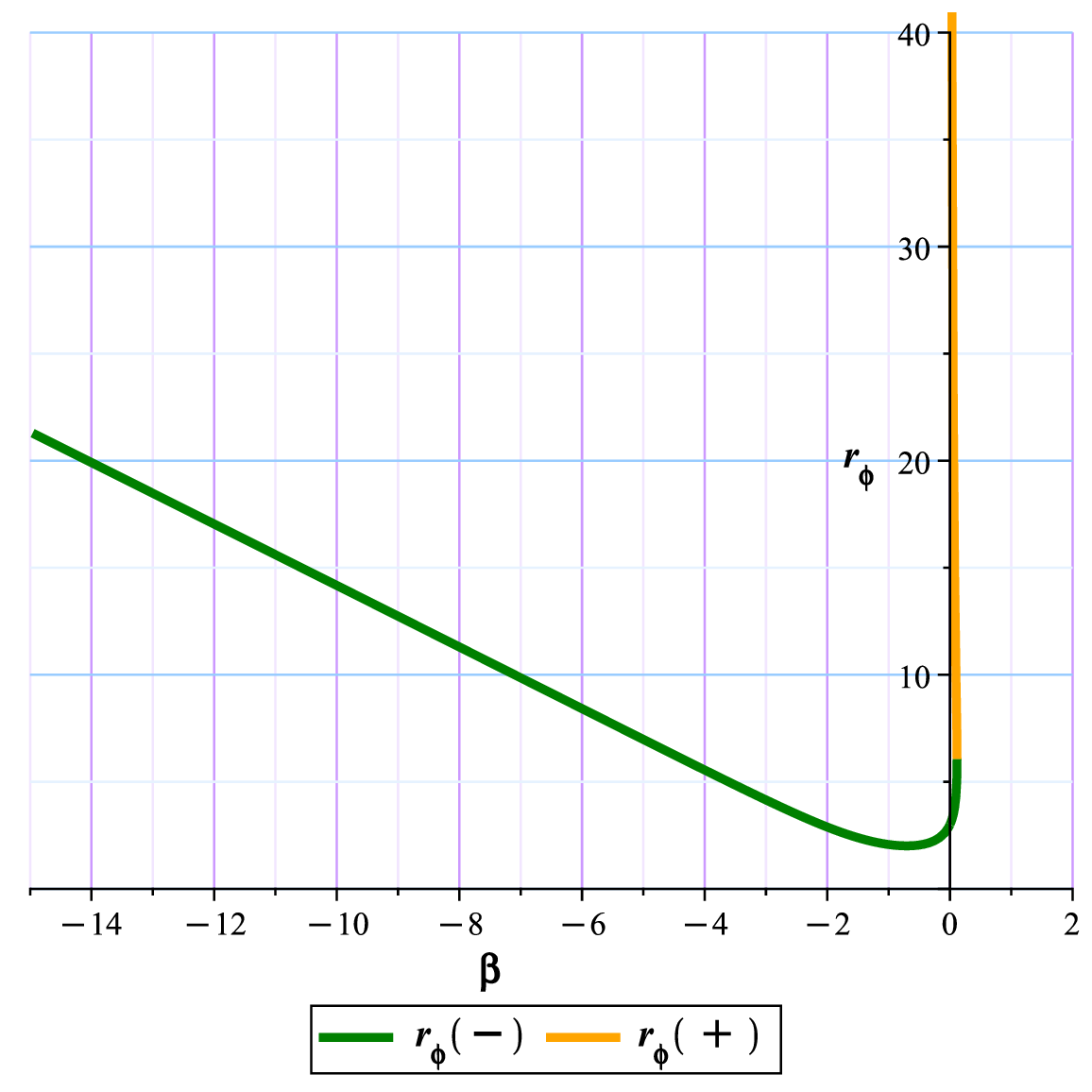}
         \caption{}
         \label{6c}
     \end{subfigure}
        \caption{\raggedright We present the topological configuration of photon spheres in BW BH. The normal vector field $n$ in the $(r-\theta)$ plane is represented by the orange arrows. The upper panels (\subref{4a}-\subref{4c}) represent the case $\alpha\leq0.05$, while the lower panels (\subref{6a}-\subref{6c}) correspond to $\alpha>0.05$; panel (\subref{4a}) and (\subref{6a}) illustrate the vector-field structure of UPS, indicated by a closed circular loop. The effective potential $H(r)$ depicted in panels (\subref{4b}) and (\subref{6b}) displays a local maximum, indicating a UPS. The allowed domain of the $\beta$ for which the radial component $\phi_{r}$, stays real are domonstrated in panels (\subref{4c}) and (\subref{6c}). The absence of a minimum in the effective potential across all the configurations reflects that stable photon spheres are forbidden by the cosmological horizon of the dS-like BW geometry.}
        \label{m4}
 \end{figure*}
 In Fig.~\ref{m4}, we graphically demonstrated the topological charge of the photon sphere Fig.~\ref{m4}(\subref{4a}), the effective potential Fig.~\ref{m4}(\subref{4b}), and the permissible values of $\beta$ for the existence of the phi function Fig.~\ref{m4}(\subref{4c}). Let's compare the permissible range for the $\beta$ of the $\phi$ function with the metric function. We observe that similar to AdS and flat cases, $\beta$ for $\phi$ function still dominates a larger range ($\beta\leq 6.593\times10^{-1}$) compared to the permissible $\beta$'s for the metric function ($\beta\leq 5.2342\times10^{-1}$). Typically, in AdS and flat models, all are part of this additional range would result in the studied structure exhibiting naked singularity behavior. However, in this case, the difference in range does not appear to yield any specific result. Because it only adds to the forbidden region, and the effective potential in this additional area is imaginary. It appears, as previously mentioned, that in this model—which ostensibly exhibits a de Sitter-like structure—the presence of a cosmological horizon effectively eliminates any permissible parameter range in which the model could exhibit naked singularity behavior. Additionally, as observed in Fig.~\ref{m3}(\subref{3b}) for this scenario and Fig.~\ref{m3}(\subref{5b}) (for case II), the effective potential $H(r)$ possesses only a single local maximum (an unstable photon sphere). Furthermore, our calculations indicate that beyond this range, the potential not only lacks any additional extrema but also becomes undefined. As a consequence, the radial component $\phi$, which represents the geometric location of photon spheres, either lacks real roots or becomes imaginary—providing strong confirmation of the model’s dS-like behavior. These results can be verified in Table~\ref{P2}.

\begin{table}[t]
\caption{\label{P2}\raggedright
Summary of the photon spheres affected by
the parameters for different cases.}
\begin{ruledtabular}
\begin{tabular}{ccccc}
   Cases & BW BH  & FP & Conditions &TTC\\
   \hline
  1 & UA & $\alpha=0.05,M=1$ & $\beta>5.234\times10^{-1} $ & nothing\\
  & UPS & $\alpha=0.05,M=1$ & $\beta\leq5.234\times10^{-1}$ &$-1$\\\hline
  2& UA & $\alpha=0.7,M=1,$ & $\beta>-1.3231 $ & nothing\\
  & UPS & $\alpha=0.7,M=1$ & $\beta\leq -1.3231$ &$-1$\\\hline
 3& UA & $\alpha=1.7,M=1,$ & $\beta> -3.36825 $ & nothing\\
  & UPS & $\alpha=1.7,M=1$ & $\beta\leq  -3.36825$ &$-1$\\
\end{tabular}
\end{ruledtabular}
\end{table}

 \subsubsection{Case II: $ 0.05 <\alpha<1 $ }
In this subsection, we examine a different range for the cosmological parameter ($\alpha$) to see if this new range can reveal different behavior for the effective potential and the topological charge of the photon sphere. Thus, for this case, we choose $m=1$ and $\alpha = 0.7$.\\
\indent In Fig.~\ref{m3}(\subref{5a}), we have graphically shown the permissible range of dark matter parameter $\beta$ and the metric function for the BW BH by substituting the $\beta=-1.5$ (green curve), $\beta=-1$ (dashed orange curve), $\beta=0.5$ (dashed red curve), and $\beta$ (dotted purple curve) in Fig.~\ref{m3}(\subref{5b}). As it is evident from Fig.~\ref{m3}, the increase of $\alpha$ significantly reduced the allowed range of $\beta\leq -1.32311$ Fig.~\ref{m3}(\subref{5a}) so that the other structure will be in the form of a BH only for negative  $\beta$ Fig.~\ref{m3}(\subref{5b}).

In this case, as in the previous case, the range of $\beta\leq 1.1557\times10^{-1}$ dominance to have the allowed $\phi^r$ is greater than the range of the metric function, but the presence of an intensified horizon prevents the formation of any potential minima (stable photon sphere) in the area and practically The total topological charges remain -1 and only UPS is allowed to appear. The results are presented in Table~\ref{P2}.

 \subsubsection{Case III: $ \alpha \geq 1 $ }
Similarly, in this subsection, we investigate the potential $H$, the photon sphere's topological charge, and the allowed values of dark matter parameter $\beta$ by using $\alpha > 1$. We have graphically presented the behavior of effective potential in Fig.~\ref{m4}(\subref{6a}) in $r - \theta$ plane, UPS with respect to the $r$ in Fig.~\ref{m4}(\subref{6b}), and the allowed range for dark matter parameter $\beta$ in Fig.~\ref{m4}(\subref{6c}). The movement process is still the same as the previous two cases; that is, the narrowing of the area dominated by $\beta$ in this case also increases with the increase of $\alpha$, the results of which can be seen in Table~\ref{P2}.\\
\indent In a comprehensive overview of the aforementioned study, it can be stated that in dS models, the presence of an enhanced horizon, referred to as the cosmological horizon, effectively prevents the formation of any potential minimum (stable photon sphere) in space. This is in stark contrast to AdS and flat models, which typically exhibit total topological charges of 0 and +1 and a stable photon sphere beyond the event horizon. Also, in this specific model, the continuous increase of the positive alpha parameter (negative values of alpha result in negative and non-physical radii) significantly reduces the domain dominated by $\beta$. 

The analysis of photon sphere stability serves a dual purpose in this study. Primarily, it acts as a geometric consistency criterion. By identifying the parameter ranges where a stable photon sphere with a topological charge of \(-1\) exists, we ensure our subsequent thermodynamic topological analysis is confined to the genuine black hole phase of the solution, characterized by a well-defined event horizon. This step is crucial to guarantee that the thermodynamic phases and critical points we investigate are intrinsic properties of the black hole spacetime, and are not inadvertently computed for a naked singularity or other non-physical configurations. In this capacity, the photon sphere analysis validates the physical domain of our main results. Secondly, this examination allows us to contextualize our findings within the causal structure of the spacetime. For the dS branch of our solution (\( \alpha^2 > 0 \)), the presence of a cosmological horizon inherently restricts the formation of certain stable photon sphere configurations. Our observation that specific thermodynamic topological charges become unattainable in this regime is, therefore, a direct reflection of this underlying causal geometry, serving as an important consistency check for our model. A more profound inquiry, inspired by recent advances in applying photon sphere methods to thermodynamic geometry, topology, and null geodesic structures \cite{6010,6011,6012}, would be to seek a direct, non-trivial correspondence between the thermodynamic topological charge \( W \) and the topological classification of photon spheres across the full parameter space. While our initial exploration does not reveal a universal, one-to-one mapping beyond the constraints imposed by the cosmological horizon in the dS case, it establishes the necessary geometric footing for our core thermodynamic investigation. The primary novelty of this work thus remains the detailed mapping of how the interplay of the parameters \( \alpha \), \( \beta \), and \( \delta \) governs thermodynamic geometry and, most distinctly, triggers global transitions in the thermodynamic topological charge \( W \), with the dark matter parameter \( \beta \) playing a decisive role.
\section{Conclusions}\label{Sec-7}
In this paper, we computed the thermodynamic variables for the BW BH and derived the mass in terms of extensive variables. We study thermodynamics, the consistency, and the divergence of the BW BH in a detailed fashion. Moreover, we have investigated the thermal stability of the BW BH in the presence of \bre~ and also determined its bound points and divergence points, in which we indeed observed the impact of cosmological parameter $\alpha$, deformation parameter $\delta$, and dark matter parameter $\beta$. In our analysis, we observed that heat capacity has a ZP $S_{0}=0.7837$, termed as a restriction point, and it is noticed that beyond this ZP $S_0$, BW BH becomes consistent (or stable). Furthermore, we have analytically and graphically demonstrated in Fig.~\ref{BDGRAPHB} and Table~\ref{table:1} that the roots and divergence points are the reducing function of dark matter parameter $\beta$, deformation parameter $\delta$, and cosmological parameter $\alpha$. Thereby, in our analysis, it is observed that cosmological parameter $\alpha$, deformation parameter $\delta$, and dark matter parameter $\beta$ have a crucial role in the consistency of BW BH.\\
\indent In addition, we utilized the thermodynamic geometry formalism to probe further the BW BH's thermodynamic conduct and phase transition with respect to \bre. For this purpose, we have employed two formalisms, which are considered fundamental formalisms. One is the Weinhold metric formalism, and the other is the Ruppeiner metric formalism. It is worth mentioning that we employed these two metric formalisms because they yielded the desired results. In the case of the Weinhold metric, it is observed that the Ricci curvature scalar, neither the divergence nor the singularity, coincides with the ZP of heat capacity, which means it did not offer us information regarding the phase transition of the BW BH.  In the case of the Ruppeiner metric, the Ricci curvature scalar's singular points coincide with the ZP of heat capacity. Thus, we inferred from our analysis that the Ruppeiner formalism provided more effective information regarding the divergence and phase transition of the BW BH as compared to the Weinhold metric. \\
\indent
This research delves into the thermodynamic topology of BW BHs, with an emphasis on detailing the mathematical framework that supports our study. This includes classifying BW BHs based on their topological properties and determining their topological charges. We have discussed these classifications' implications and compared topological thermodynamics' properties computed from the Bekenstein-Hawking and \bre, highlighting key differences and similarities. Our investigation emphasizes the critical role of the parameter $\beta$ in determining the number of topological charges. When $\delta$ and $\alpha$ are held constant, changes in $\beta$, it leads to topological charge $(\omega = 0)$ and $(\omega = -1)$. Interestingly, changes in $\delta$ or $\alpha$ also somehow affect the number of topological charges, and also underscore the dominant role of $\beta$. Similar outcomes are observed when $\delta$ is zero, emphasizing $\beta$'s impact. Therefore, thermodynamic instabilities arise only when \bre~ is employed, while remaining absent under Bekenstein-Hawking entropy, and the dark matter parameter plays a leading role in determining the topological classes, revealing characteristics that are not straightforward consequences of earlier models. These results also distinguish between behaviors that are universal and those that are inherent to the specific BW BH examined in this study. \\
\indent In the context of the dS model, the cosmological horizon effectively inhibits the formation of potentially stable photon spheres. As a result, achieving topological charges of 0 and +1 is not feasible within this framework. Furthermore, we examined how gradual adjustments to the model's alpha parameter lead to a reduction in the $\beta$-dominated region. These findings significantly impact our understanding of BH thermodynamics and topology. The dominant influence of the parameter $\beta$ suggests that it could be a key factor in future theoretical models and simulations. Additionally, the differences observed between dS, AdS, and flat models highlight the importance of the cosmological context in BH studies.\\
\indent Future research could further investigate the interplay between different parameters and their impact on topological charges, potentially uncovering new insights into the nature of BHs. Moreover, extending this analysis to other types of BHs and higher-dimensional models could provide a more comprehensive understanding of these fascinating objects' thermodynamic and topological properties. In conclusion, our study underscores the critical role of the parameter $\beta$ in the thermodynamic topology of BW BHs and opens up new avenues for research in this intriguing field. The insights gained from this work could pave the way for more detailed and nuanced models of BH behavior, enhancing our overall comprehension of these enigmatic cosmic entities.
\section*{Acknowledgments}
The work of KB was supported by the JSPS KAKENHI Grant Numbers JP21K03547, 24KF0100, 25KF0176, and Competitive Research Funds for Fukushima University Faculty (25RK011). We would like to sincerely thank Dr. Krishnakanta Bhattacharya for the constructive and insightful discussion. 

\section*{Appendix A: Additional figures for the allowed ranges}\label{Sec:App_1}
\begin{figure*}
     \centering
     \begin{subfigure}[b]{0.35\textwidth}
         \centering
         \includegraphics[width=\textwidth]{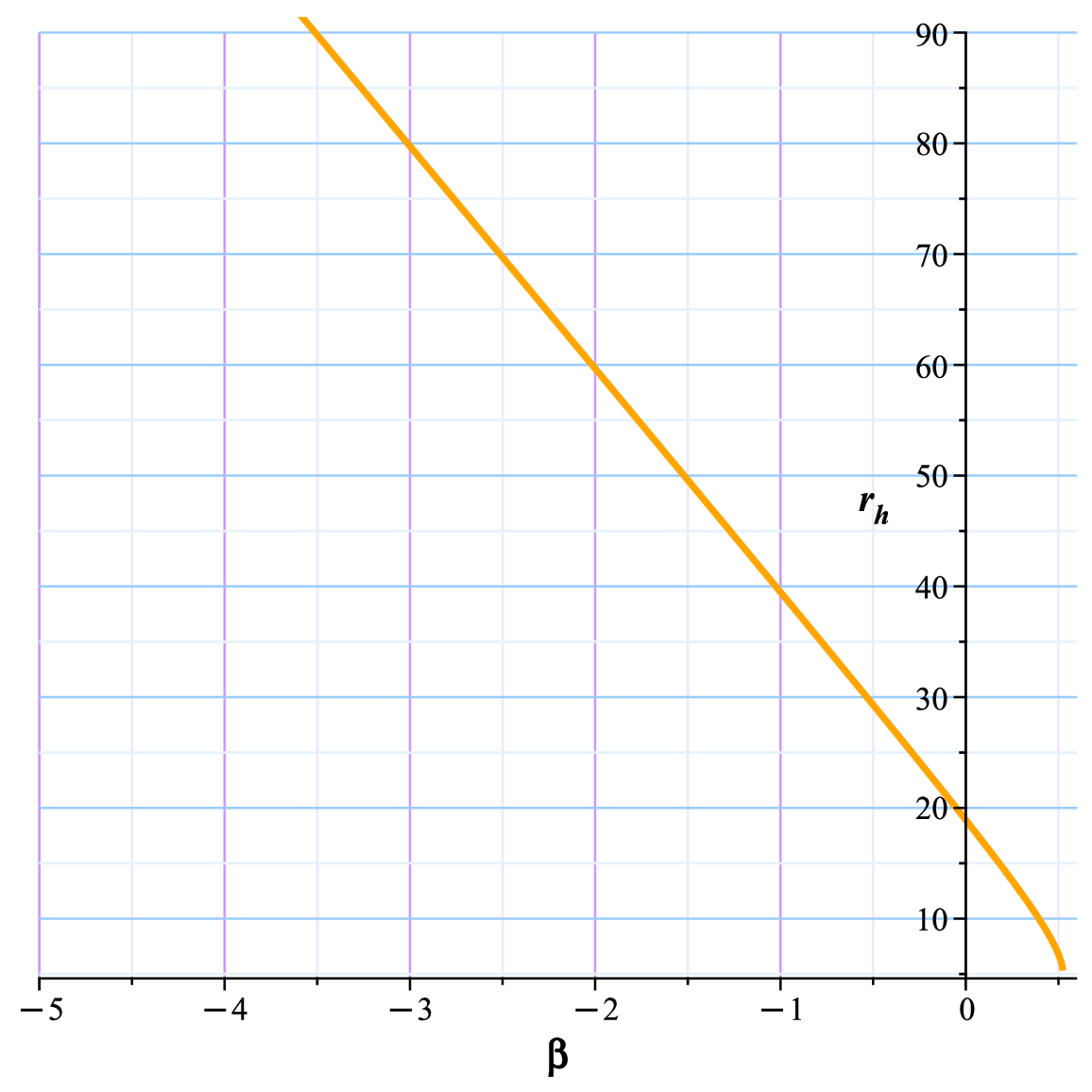}
         \caption{}
         \label{3a}
     \end{subfigure}
     \begin{subfigure}[b]{0.35\textwidth}
         \centering
         \includegraphics[width=\textwidth]{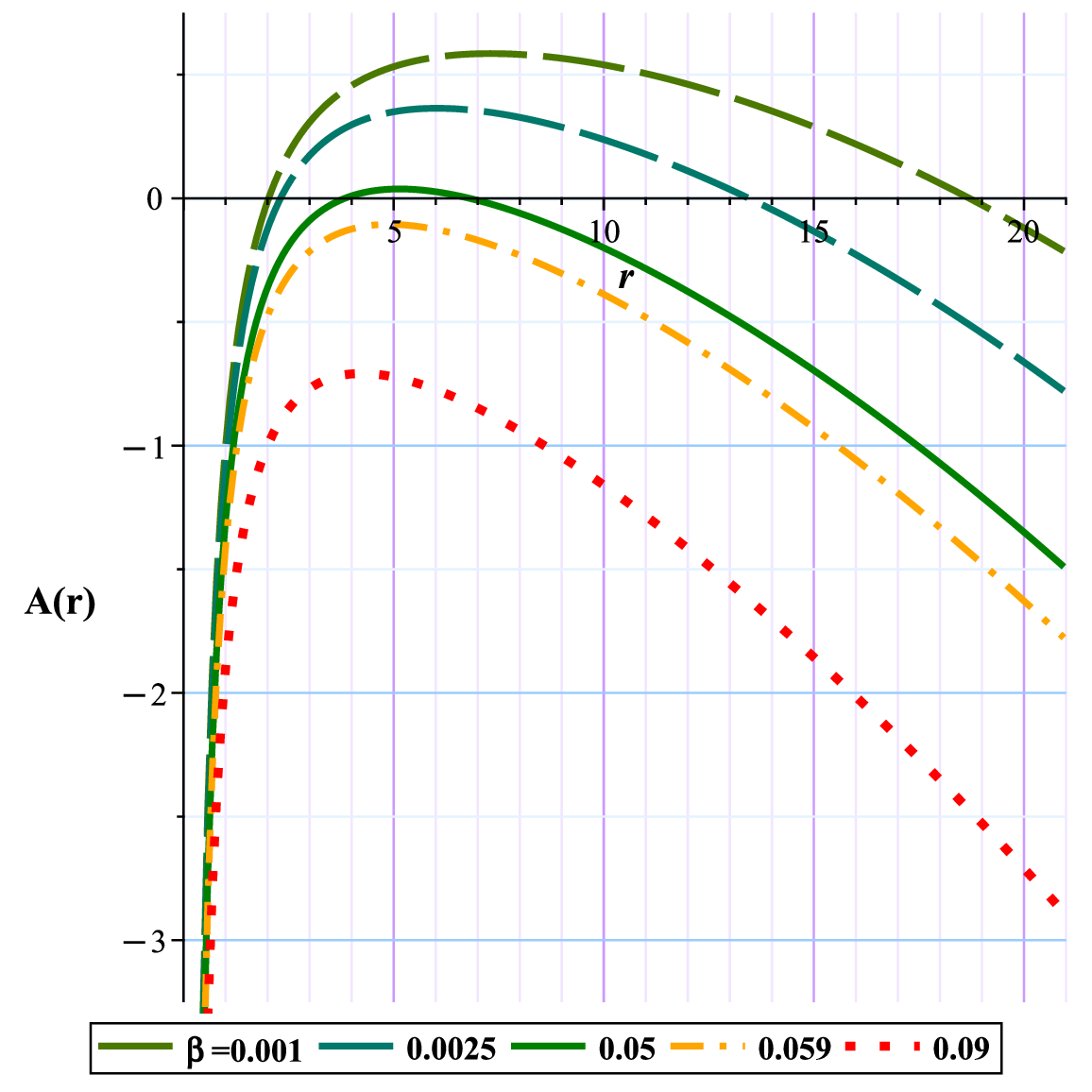}
         \caption{}
         \label{3b}
     \end{subfigure}
     \begin{subfigure}[b]{0.4\textwidth}
         \centering
         \includegraphics[width=\textwidth]{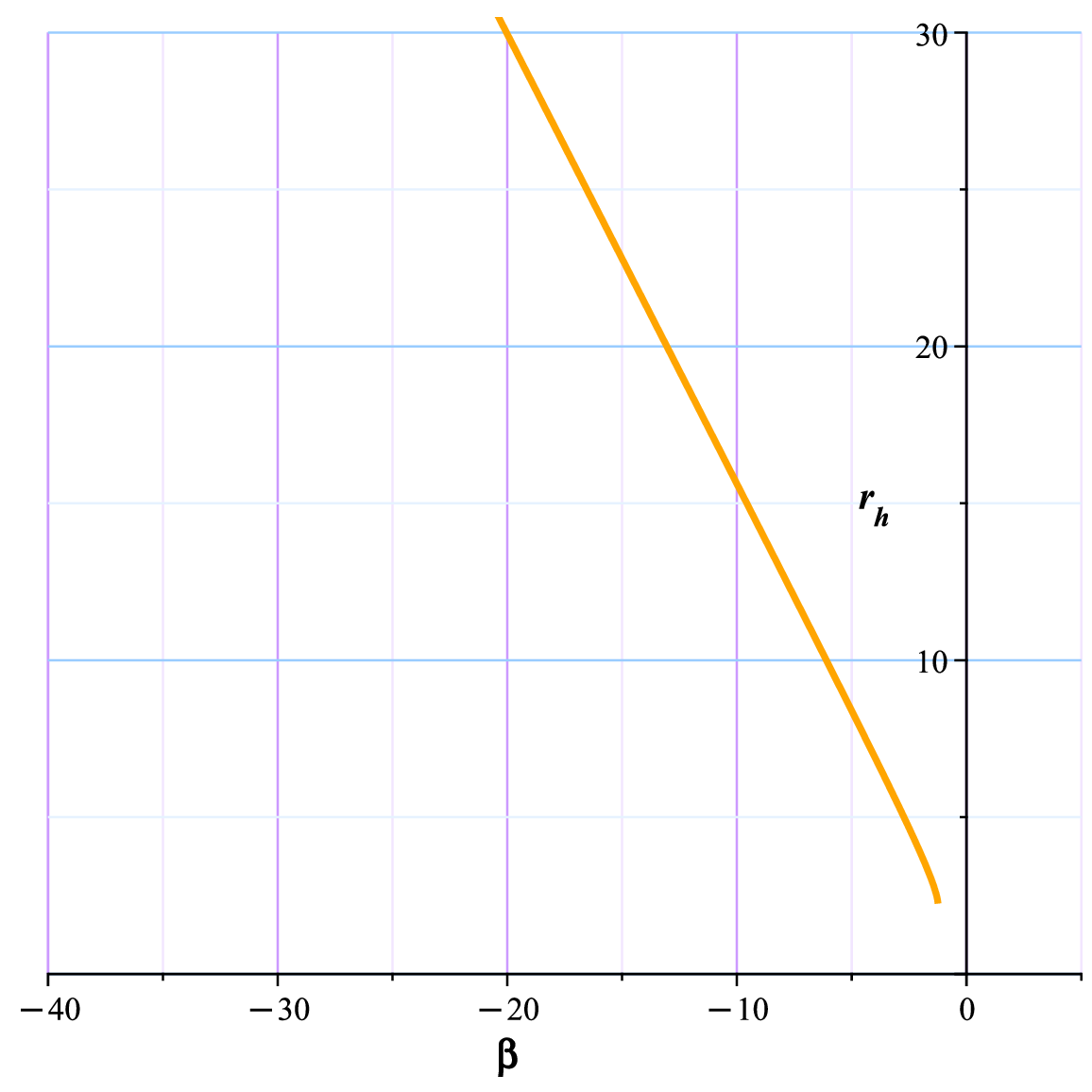}
         \caption{}
         \label{5a}
     \end{subfigure}
     \begin{subfigure}[b]{0.4\textwidth}
         \centering
         \includegraphics[width=\textwidth]{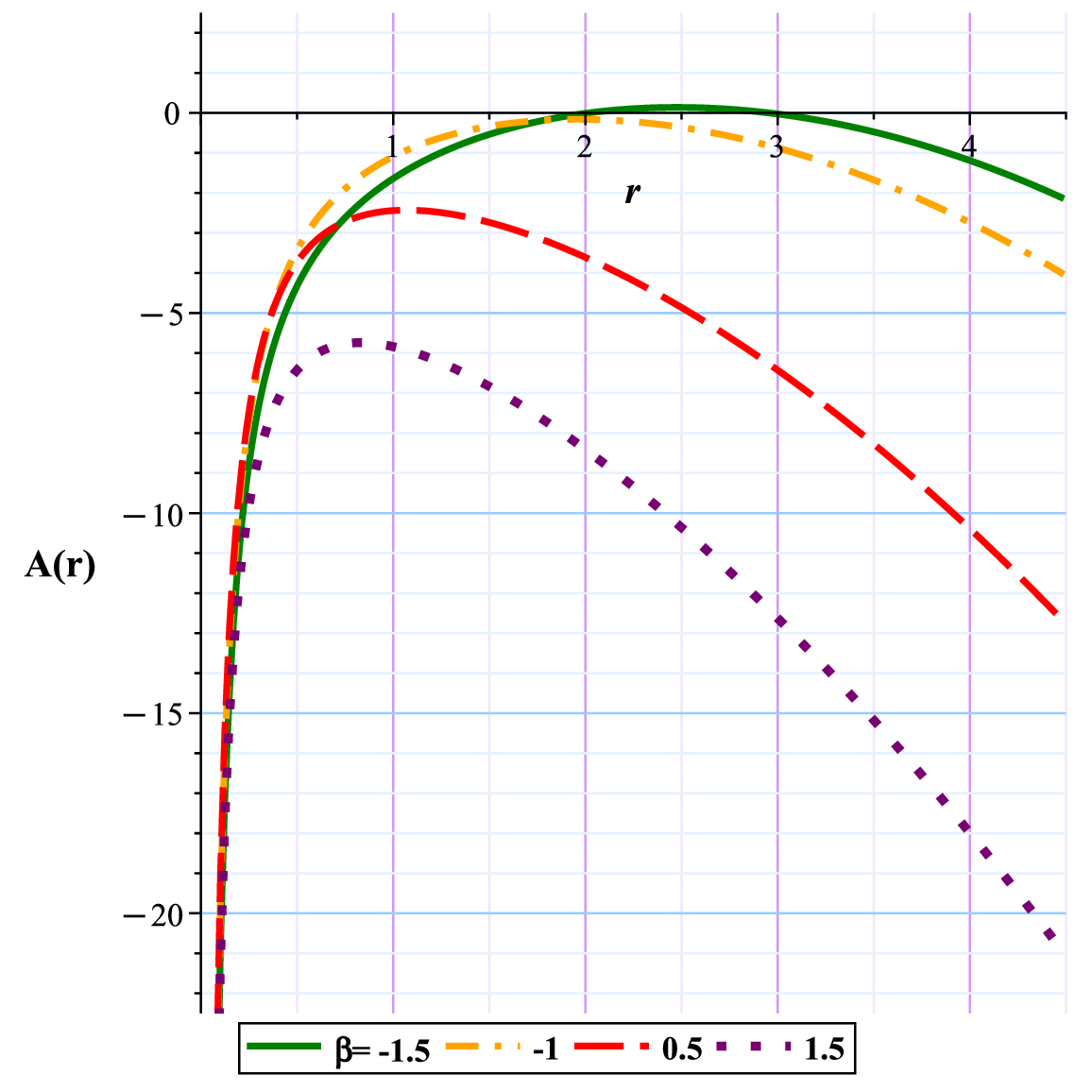}
         \caption{}
         \label{5b}
     \end{subfigure}
        \caption{\raggedright Figs.~\ref{m3}(\subref{3a}) and \ref{m3}(\subref{5a}): The allowed range of $\beta$ to have a positive real horizon in BW BH. Figs.~\ref{m3}(\subref{3b}) and \ref{m3}(\subref{5b}): Metric function with different $\beta$ for BW BH. In Fig.~\ref{m3}(\subref{3b}), we have presented the metric function by putting the dark matter parameter $\beta=0.001$ (green curve), $\beta=0.0025$ (light blue dashed curve), $\beta=0.05$ (dashed green curve), $\beta=0.059$ (dotted dashed orange curve), and  $\beta=0.09$ (dotted red curve). In Fig.~\ref{m3}(\subref{5b}), we have plotted the metric function by using $\beta=-1.5$ (green curve), $\beta=-1$ (dotted dashed orange curve), $\beta=0.5$ (dashed red curve), and $\beta=1.5$ (dotted purple curve).}
        \label{m3}
\end{figure*}

In this appendix, we present the graphical behavior of the metric function to explicitly describe the allowed range for the dark matter parameter $\beta$ to have positive values in the BW BH for both cases $\alpha\leq0.05$ and $\alpha>0.05$ in Figs.~\ref{m3} upper and lower panels, respectively. Furthermore, let us mention that we discuss the behavior of these figures in Sec.~\ref{Sec-6}.

\end{document}